%% file: main.tex
\documentclass[sigconf, anonymous=false]{acmart}

\input{macros}

\AtBeginDocument{%
  \providecommand\BibTeX{{%
    \normalfont B\kern-0.5em{\scshape i\kern-0.25em b}\kern-0.8em\TeX}}}

\setcopyright{acmcopyright}
\copyrightyear{2023}
\acmYear{2023}
\acmDOI{XXXXXXX.XXXXXXX}

\acmConference[KDD ’24]{ACM KDD Conference 2024}{August 25-29, 2024}{Barcelona, Spain}
\acmPrice{15.00}
\acmISBN{978-1-4503-XXXX-X/18/06}

\begin{document}

\title{Understanding team collapse via probabilistic graphical models}

\author{Iasonas Nikolaou}
\affiliation{%
  \institution{Boston University}
  \city{Boston}
  \country{USA}
}
\email{nikolaou@bu.edu}

\author{Konstantinos Pelechrinis}
\affiliation{%
  \institution{University of Pittsburgh}
  \city{Pittsburgh}
  \country{USA}
}
\email{kpele@pitt.edu}

\author{Evimaria Terzi}
\affiliation{%
  \institution{Boston University}
  \city{Boston}
  \country{USA}
}
\email{evimaria@bu.edu}

\begin{abstract}
 In this work, we develop a graphical model to capture team dynamics. We analyze the model and show how to learn its parameters from data. Using our model we study the phenomenon of \emph{team collapse} from a computational perspective. We use simulations and real-world experiments to find the main causes of team collapse. We also provide the principles of building resilient teams, {\ie}, teams that avoid collapsing. Finally, we use our model to analyze the structure of NBA teams and dive deeper into games of interest. 
 \end{abstract}

\ccsdesc[300]{Information systems~Data mining}
\ccsdesc[500]{Computing methodologies~Learning in probabilistic graphical models}
\ccsdesc[500]{Theory of computation~Social networks}
\keywords{team collapse, team dynamics, probabilistic graphical models}

\maketitle

\import{sections/}{intro}



\import{sections/}{model}


\import{sections/}{simulations}

\import{sections/}{learning}



\import{sections/}{experiments}

\import{sections/}{conclusions}

\clearpage

\bibliographystyle{ACM-Reference-Format}
\bibliography{references}


\newpage
\onecolumn
\appendix
\import{sections/}{appendix}

\end{document}

%% file: macros.tex
\usepackage{import}
\usepackage{amsmath}
\usepackage{amsthm}
\usepackage{multirow}
\usepackage{url}
\usepackage{tikz}
\usepackage{subcaption}

\newcommand{\figwidth}{0.48\textwidth}
\newcommand{\PBP}{\texttt{PBP}}
\newcommand{\NBAseason}{\texttt{NBAseason}}
\newcommand{\NBAgames}{\texttt{NBAgames}}

\DeclareMathOperator*{\argmax}{arg\,max}

\newcommand{\vect}[1]{\mathbf{#1}}

\newcommand{\Prob}{\text{Pr}}
\newcommand{\params}{\boldsymbol{\Theta}}

\newcommand{\PIE}{\texttt{PIE}}

\newcommand{\Ob}{\ensuremath{\mathbf{O}}}
\newcommand{\Hb}{\ensuremath{\mathbf{H}}}

\newcommand{\h}{\ensuremath{\mathbf{h}}}

\newcommand{\I}{\ensuremath{\text{Infl}}}

\newcommand{\EX}{\ensuremath{\mathbb{E}}}

\newcommand{\Mmatrix}{\ensuremath{\mathcal{M}}}
\newcommand{\Nmatrix}{\ensuremath{\mathcal{N}}}
\newcommand{\Rmatrix}{\ensuremath{\mathcal{R}}}
\newcommand{\Ematrix}{\ensuremath{\mathcal{E}}}

\newtheorem{lemma}{Lemma}
\newtheorem{definition}{Definition}

\newtheorem{remark}{Remark}

\newcommand{\spara}[1]{\smallbreak\noindent{\bf{#1}}}
\newcommand{\mpara}[1]{\medskip\noindent{\bf{#1}}}

\newcommand{\etal}{\textit{et al.}}
\newcommand{\ie}{\textit{i.e.}}
\newcommand{\eg}{\textit{e.g.}}

\usepackage[show]{chato-notes}
\usepackage{algorithm}
\usepackage{algpseudocode}

\newcommand*{\belowrulesepcolor}[1]{%
	\noalign{%
		\kern-\belowrulesep
		\begingroup
		\color{#1}%
		\hrule height\belowrulesep
		\endgroup
	}%
}
\newcommand*{\aboverulesepcolor}[1]{%
	\noalign{%
		\begingroup
		\color{#1}%
		\hrule height\aboverulesep
		\endgroup
		\kern-\aboverulesep
	}%
}

\newcommand{\squishlist}{\begin{list}{$\bullet$}
  { \setlength{\itemsep}{0pt}
     \setlength{\parsep}{3pt}
     \setlength{\topsep}{3pt}
     \setlength{\partopsep}{0pt}
     \setlength{\leftmargin}{1.5em}
     \setlength{\labelwidth}{1em}
     \setlength{\labelsep}{0.5em} } }
\newcommand{\squishend}{
  \end{list}  }

%% file: sections/intro.tex
\section{Introduction}
What are the reasons that a (sports) team severely under-performs and eventually \emph{collapses}?
Are some teams more prone to collapse than others? 
Are there some guiding principles for building resilient teams?

In this paper, we make a significant step towards addressing these questions. We achieve this by introducing a probabilistic graphical model that assumes the following:
at any time $t$ a player's \emph{hidden} (mental) state depends on the 
player's own \emph{hidden} state at $t-1$ and the 
observed performances of all team members at $t-1$.
 Consequently, a \emph{team collapse} happens at time $t$ if more than a certain percentage of
the team's players are in a ``low" hidden (mental) state at $t$.

Our model parameters enable us to encode different types of dependencies among team members. For example, there may be 
players whose performance is only affected by their own hidden/observed state
and is unaffected by the performance of others. We call such players \emph{pillars}. On the other hand, non-pillar players are more susceptible to be influenced by their teammates' observed performance.
An interesting question we explore is to what extend having more or less pillars affects the resilience of a team and influences its collapse probability. Moreover, is it just the number of pillars important or their dependency patterns as well?
We believe that being able to answer questions like this 
can inform 
roaster-building decisions.

From the computational point of view, we show that we can use the
Expectation-Maximization (EM) framework to learn the parameters of our model.  In particular, we borrow several ideas from probabilistic graphical models~\cite{koller2009probabilistic}
and Hidden Markov Models (HMMs)~\cite{bishop06pattern} in order to implement the expectation and the maximization
steps of EM. 
In practice, we show that despite the large number of parameters of our model, EM can learn those parameters effectively in a small number of iterations.

Using real data from National Basketball Association (NBA) games (season 2021-22), we deploy our EM algorithm to learn the parameters 
of our model for all NBA teams. We identify some interesting
dependencies among players and demonstrate that pillar players do not 
always coincide with players widely acknowledged as ``stars".
Moreover, our framework allows us to identify team collapses in various games and further analyze them in order to find the reasons for which these collapses happened. We believe that these results 
provide useful insights on team dynamics and can be used
from team psychologists and General Managers in order to safeguard teams from collapses.

Overall, our model  allows us
to capture complex team dynamics and can be used as a tool 
by experts in team sports to analyze 
their teams' compositions, simulate the performance of 
different compositions and use these insights to 
create teams that are more resilient to team collapses. 

Although related to other probabilistic models proposed in the literature (see related work for an extended discussion), the exact details of our model and the corresponding computational problem associated with the learning of its parameters are novel. 
This is because, to the best of our knowledge, our model is the first probabilistic graphical model that
is tailored to capture how teammate dependencies give rise to team collapses.

\mpara{Related work:}
During the last decade, the ability to collect multiple player-tracking data has led to a growth of computational approaches for evaluating players, teams and
team strategies in a variety of sports~\cite{cervone2016multiresolution,chen2021instant,damour15,decroos2019actions,miller14,nguyen2023here,lucey2014quality,seidl2018bhostgusters,sicilia2019deephoops,yue2014learning,yurko2020going}.
From the application point of view, our work falls into this general area and it relates to work
on sport analytics and athlete choking. 
From the computational point of view, our work is related to existing work on probabilistic models and interconnected Hidden Markov Models (HMMs). We summarize these connections below.

\paragraph{Team and player performance ratings:} One of the main tasks that the field of sports analytics has dealt with is that of team and player performance evaluation. 
This typically is achieved through models that learn team and player ratings, which are then used in 
 match prediction and gambling or to help teams identify players to acquire.  
The simplest team ratings are based on regression methods
~\cite{langville2012s,w2007quantile,bradley1952rank,winston2022mathletics}. More sophisticated ones use network analysis of the  
``who-wins-whom'' network 
~\cite{park2005network,pelechrinis2016sportsnetrank}, while others use latent space models to learn an embedding for each team that will help obtain a team ranking \cite{chen2016modeling,chen2016predicting,pelechrinis2018linnet}. 
Similarly, for evaluating players the traditional approach for obtaining player ratings is using regression based approaches \cite{macdonald2011regression,thomas2013competing,sabin2021estimating} to model the score differential during a game possession as a function of the players on the court/pitch. 

These ratings ignore dependencies between the players and make the implicit assumptions that player ratings are additive, {\ie}, a team is the sum of the individual players. 
As such, it is hard for these type of models to help us understand and explain situations where a team significantly over or under performs its expectation within a game. 
In our work, we explicitly consider the  dependencies between players and we develop a framework that can explain how these dependencies can lead to a team collapsing. 

\paragraph{Athlete's ``choking''}: The common belief is that super-star athletes shine when the stakes are high, while athletes in lower tiers severely under perform under pressure, {\ie}, they ``choke'' or collapse. 
While there has been some literature on players performance and decision-making under pressure, this has traditionally been in the field of sports psychology \cite{baumeister1986review,pineda2022play,parkin2017gunslingers}. 
The vast majority of the literature has been on the {\em complementary} phenomenon of ``clutch'' players, {\ie}, players that deliver when the stakes are high. 
The majority of this literature indicates that individual clutch performance is highly noisy \cite{solomonov2015clutch,wallace2013homo,birnbaum2008clutch}. 
One of the few computational studies on player performance under pressure is that from Bransen {\etal}~\cite{bransen2019choke} who develop  metrics to evaluate the performance of individual soccer players under situations with high mental pressure. 
In a similar direction, Dohmen \cite{dohmen2008professionals} focused on studying penalty shootouts and found that players of the home team are more likely to choke; the same study identified no evidence in their data that high stakes induce choking. 
A detailed overview of mechanisms and potential moderators of choking in sport is further provided by Hill {\etal}~\cite{hill2010choking}. 

All the above work considers individual players' choking.
Our work enhances this literature by studying the {\em team effects} of individual performance, as well as, the possible influence they exert on the performance of their teammates and how this translates to a team's ability to withstand, or not, periods of subpar performance of individual players.


\paragraph{Computational work on interacting HMMs}
Our model is inspired by HMMs~\cite{koller2009probabilistic}
and in particular work on ``coupled" HMMs~\cite{dong12graphcoupled,pan2012modeling} where
the nodes of a network are HMMs themselves influenced by the other
nodes via the edges of the network. The most related to ours is the work of Pan {\etal}~\cite{pan2012modeling}. Motivated by applications of influence in social media and disease propagation, Pan {\etal} propose a model 
very similar to ours, where the hidden state of every individual 
at time step $t$ depends on their own hidden state at $t-1$ and 
the others' hidden states at $t-1$.  Contrary to this, our model assumes that an individual's hidden state at $t$ depends on their own
hidden state at $t-1$ and the others' \emph{observed} states at $t-1$.
From the application point of view this difference is important: players in a team don't know the hidden state of their teammates; they only know their teammates' observed performance. Thus, our model is better suited for our application. The computational consequences of this
difference are also important: we are able to 
compute (and sample) from the posterior probability distribution during
the E-step of our EM algorithm, while they cannot as they have to deploy variational inference techniques.

\import{img/}{graphical-model}

%% file: img/graphical-model.tex
\begin{figure}
    \centering

    \begin{tikzpicture}[thick, scale=0.9]
      \node[circle, draw] (A) at (0, 2) {$H_1^1$};
      \node[circle, draw] (B) at (0, 0) {$H_1^2$};
      \node[circle, draw, fill=gray!30] (C) at (1.5, 0.5) {$O_1^2$};
      \node[circle, draw, fill=gray!30] (D) at (1.5, 1.5) {$O_1^1$};
      \node[circle, draw] (E) at (3, 2) {$H_2^1$};
      \node[circle, draw] (F) at (3, 0) {$H_2^2$};
      \node[circle, draw, fill=gray!30] (G) at (4.5, 0.5) {$O_2^2$};
      \node[circle, draw, fill=gray!30] (H) at (4.5, 1.5) {$O_2^1$};
      \node[circle, draw] (I) at (6, 2) {$H_3^1$};
      \node[circle, draw] (J) at (6, 0) {$H_3^2$};
      \node[circle, draw, fill=gray!30] (K) at (7.5, 0.5) {$O_3^2$};
      \node[circle, draw, fill=gray!30] (L) at (7.5, 1.5) {$O_3^1$};
      
      \draw[->] (A) edge[bend left] (E);
      \draw[->] (B) edge[bend right] (F);
      \draw[->] (A) -- (D);
      \draw[->] (B) -- (C);
    
      \draw[->] (D) -- (E);
      \draw[->] (D) -- (F);
      \draw[->] (C) -- (E);
      \draw[->] (C) -- (F);
      
      \draw[->] (E) edge[bend left] (I);
      \draw[->] (F) edge[bend right] (J);
      \draw[->] (E) -- (H);
      \draw[->] (F) -- (G);
    
      \draw[->] (H) -- (I);
      \draw[->] (H) -- (J);
      \draw[->] (G) -- (I);
      \draw[->] (G) -- (J);
      \draw[->] (I) -- (L);
      \draw[->] (J) -- (K);
    \end{tikzpicture}
    \caption{Graphical model representation of our model for $n=2$ entities and $T=3$ timesteps}
    \label{fig: graphical_model}
\end{figure}
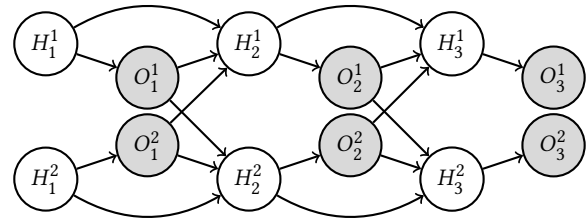

%% file: sections/model.tex
\section{The model}\label{sec:model}
We start this section by giving some notational conventions we follow throughout the paper. Then, we give a high-level description of our model, and provide example instantiations that enable us to better convey its expressivity. 

\subsection{Notational conventions}
\input{sections/notation}

\subsection{Model description}
Our team model aims to capture the influence that the team entities/players
have on each other. The model is inspired by Hidden Markov Models and their interaction as described by Pan {\etal}\cite{pan2012modeling}. 
Despite this initial inspiration, our model deviates from the aforementioned work in some key ways which have significant application and computational consequences, discussed in the last paragraph of the Related work.

\spara{Entities:}
The model assumes of $n$ entities. 
At time step $t=1,\ldots , T$, each entity $i$ is associated with a \emph{hidden} (not observed) state $H_t^i \in \mathcal{H}$, where $\mathcal{H}$ is a discrete finite set of hidden states with cardinality $n_h$. 
At $t$, entity $i$ emits an \emph{observed} signal $O_t^i \in \mathcal{O}$, where $\mathcal{O}$ is a discrete finite set of \emph{observed states} with cardinality $n_o$. We assume that we have observations for $T$ timesteps.

In our case, the entities correspond to the \emph{players}, members of a team.
Since we consider basketball teams we have $n=5$. 

The hidden states can be thought of as the psychological/mental states of each player. 
In our experiments, we define $3$ hidden states $\mathcal{H} = \{-1, 0, 1\}$ where $-1$ is the lowest (worst), $0$ is the average and $3$ is the highest (best) possible mental state of a player.
The observed states capture the performance of each player during a game. 
We again experiment with $3$ observed states  $\mathcal{O}= \{-1, 0, 1\}$;
 $0$ corresponds to a player performing within their average performance, $1$ (resp.\ $-1$) corresponds to 
 a player's performance above (resp.\ below) his/her expected performance. 
We give the details of how we set those in real data in Section~\ref{sec: real-world experiments}.

\spara{Model parameters:}
The key assumption of our model is that the hidden state $H_t^i$ of an entity $i$ at time $t$ depends on the observed states of all entities at time $t-1$, denoted by 
$\Ob_{t-1}$, as well as $H_{t-1}^i$ ({\ie}, the hidden state of $i$ at time $t-1$).
Figure~\ref{fig: graphical_model} shows the graphical model representation of our model for $n=3$
and $T=3$.

Central for our model is the computation of the conditional probability
$$
\Prob(H_t^i\mid \Ob_{t-1}, H_{t-1}^i).
$$
In general, computing this probability is hard due to the exponential state space.   
Thus, we adopt the following simplifying assumption:
\begin{eqnarray} \label{eq: cond-prob} 
\lefteqn{
\Prob(H_t^i\mid \Ob_{t-1}, H_{t-1}^i) = }\\
    & & \sum_{j=1}^n \underbrace{R_{ij}}_{\text{tie strength}} \times
    \underbrace{\I(H_t^i \mid O_{t-1}^j)}_{\text{influence } j \rightarrow i}
    \; + 
    \underbrace{r_i \times \I(H_t^i \mid H_{t-1}^i)}_{\text{previous hidden state influence}},\nonumber
\end{eqnarray}
where $R_{ij}$'s and $r_{i}$ are convex coefficients; {\ie},
$ \sum_{j=1}^n R_{ij} + r_i = 1$ and   $R_{ij}, r_{i} \geq 0$.
For brevity, we also use vector $\Rmatrix^i = (r_i\mid R_{i1}, \dots, R_{in})$ to store the 
convex coefficients that appear in Eq.~\eqref{eq: cond-prob}.
Finally, we use $\Rmatrix$ to denote the $n\times (n+1)$
matrix that contains all $n$ $\Rmatrix^i$ vectors as its rows.

Note that $\Rmatrix$ is an important parameter as it encodes the structure of influence within the team. In Section~\ref{sec: real-world experiments} this is the matrix we are inspecting in order to analyze the structure of the teams we consider.  We refer to  $\Rmatrix$  as the \emph{team structure}.

$\I(H_t^i = h \mid O_{t-1}^j = o)$ corresponds to the influence that
$O_{t-1}^j=o$ has on $H_{t}^i=h$.  We 
store these influence values in $ n_o\times n_h$
matrices $\Mmatrix^{ij}=\I(H_t^i \mid O_{t-1}^j)$.
Similarly, $\I(H_t^i = h \mid H_{t-1}^i =h')$ corresponds to the influence that $H_{t-1}^i=h'$ has
on $H_{t}^i=h$. For notational convenience, we define an $n_h\times n_h$
matrix $\Nmatrix^{i}$ storing these influence values; {\ie},  $\Nmatrix^{i} =\I(H_t^i \mid H_{t-1}^i)$. We assume that both matrices $\Mmatrix^{ij}$ and $\Nmatrix^{i}$ are row-stochastic. 

One may be tempted to interpret these influence values as probabilities
and assume that $\I(H_t^i \mid O_{t-1}^j) = \Prob(H_t^i \mid O_{t-1}^j)$ and
$\I(H_t^i \mid H_{t-1}^i)=\Prob(H_t^i \mid H_{t-1}^i)$. In Appendix~\ref{app:influence_matrices}
we show that this is not true and that the $\I (\cdot)$ values are not necessarily probabilities.
In order to establish that our model is well-defined we show 
the following lemma, with proof in Appendix~\ref{app:problaw}:
\begin{lemma}\label{lemma:model_probability_law}
$\Prob(\cdot \mid \Ob_{t-1}, H_{t-1}^i)$
 induces a probability law.
\end{lemma}

Finally, we need to specify the \emph{emission probabilities}
 $\Prob(O_t^i | H_t^i)$ that connect the hidden state of an entity to its
 observed state. We store those in matrix:
 $\Ematrix^i = \Prob(O_t^i | H_t^i)$.

 Matrices $\Mmatrix^{ij}$, $\Nmatrix^i$, $\Rmatrix$ and $\Ematrix^i$ are \emph{the parameters}
 of our model denoted by $\params = \{\Mmatrix^{ij}, \Nmatrix^i, \Rmatrix,\Ematrix^i\mid i,j\in [n] \}$.
These parameters fully describe our model and each instantiation of the parameters corresponds to 
a different \emph{team profile}.

In many practical applications it is reasonable to assume that matrices $\Mmatrix^{ij}$ are the same for all $j \in [n]$. In such cases we write $\Mmatrix^i$, instead of $\Mmatrix^{ij}$.
Simplifying the model even further we may have all entities to share the same $\Mmatrix$ matrix. 
In practice, it is also reasonable to assume that there is a single $\Nmatrix$, and $\Ematrix$ shared among all entities.
In our experiments we make such simplifying assumptions assuming that its parameters
are $\params = \{\Mmatrix, \Nmatrix, \Rmatrix,\Ematrix \}$.

\subsection{Team collapse}
Given the above model, we can now formally define the \emph{hidden} and the \emph{observed team states} and the notions of
\emph{player choking} and \emph{team collapse}.

\begin{definition}\label{dfn:team_state}
The \emph{hidden (resp.\ observed) team state} is the average hidden (resp.\ observed) state
of the team's players. 
\end{definition}

\begin{definition}\label{dfn:choking}
    A player $i$ is in a \emph{chocking state}
    at time $t$, if $H_t^i = -1$, \ie, the player is in the lowest possible hidden state.
\end{definition}

\begin{definition}\label{dfn:collapse}
A team is in an $\alpha$-\emph{collapse state} at time $t$, if 
   $\alpha$ fraction of the team's players are in a \emph{chocking} state at time $t$; for $\alpha\in[0,1]$. 
\end{definition}

We use these definitions of player choking and team collapse when analysing the expressivity of our model in Section~\ref{sec:expressivity} and when doing experiments with data from NBA teams in Section~\ref{sec: real-world experiments}.

%% file: sections/notation.tex
For the rest of the paper, we establish the following notation: we use uppercase letters to denote random variables, e.g. $X$, $O$, $H$ and lowercase letters to denote non-random quantities, e.g. $x$, $o$, $h$.  When the random variables are vectors we denote them with bold uppercase letters
$\vect{X}$, $\vect{O}$ and $\vect{H}$ and the corresponding vector non-random quantities
$\vect{x}$, $\vect{o}$ and $\vect{h}$.
For a vector random (resp.\ non-random) variable
$\vect{X}$ (resp.\ $\vect{x}$), we denote its $i$-th coordinate 
with $X^i$ (resp.\ ${x}^i)$.
We also use uppercase calligraphic letters to denote (non-random) matrices, e.g. $\mathcal{M}, \mathcal{N}$.
All matrices in this paper are  non-random quantities.  In some cases, we abuse notation, and we also use
use uppercase calligraphic letters to denote sets, e.g. $\mathcal{H}$, and $|\cdot|$ to denote their cardinality. 
We use $\Prob(X=x)$ to denote the probability of random variable $X$ taking the value $x$. 
For brevity, when clear from context, we abuse the notation and use $\Prob(X)$ instead of $\Prob(X=x)$. 

%% file: sections/simulations.tex
\section{Exploring team profiles}\label{sec:expressivity}
In this section, we experimentally evaluate how prone are different team profiles
to team collapse. For this, we focus on a few basic team profiles, {\ie}, different
instantiations of the model parameters $\Theta = \{\Mmatrix, \Nmatrix, \Rmatrix, \Ematrix\}$\footnote{We focus on the simplified version of the model.}.
Then, we generate data using the instances of $\Theta$ we consider and study different properties of the corresponding team profile. These properties include the \emph{average hidden (resp.\ observed) team state} (Def.~\ref{dfn:team_state}) and the probability that a team gets into a collapse state (Def.~\ref{dfn:collapse}).

The goal of this section is not only to study the properties of the particular profiles
we are considering, but also demonstrate how our model can be used by experts 
in team sports to analyze different team compositions and their robustness to team collapses. 

\subsection{Team profiles}
Below we describe the design principles for the team profiles we experiment with in the main body of the paper; a more extensive list of team profiles as well as some specific instances of the parameters are also shown in Appendix~\ref{app:expressivity}.

\spara{Setting $\Mmatrix$:}  
Matrix $\Mmatrix$ captures the influence of everyone's observed states to the hidden state of a player. When designing $\Mmatrix$, we assume that   
 $\Mmatrix(o=\ell, h = \ell)>\Mmatrix (o=\ell,h = \ell')$ for all $\ell\neq\ell'$ and $\ell, \ell'\in\{-1,0,1\}$. 
 That is, when observing a state that is similar to their hidden state, the players do not change their hidden state. Moreover,
$\Mmatrix(o=\ell, h = \ell')$ is monotonically decreasing with the difference
$|\ell - \ell'|$. Intuitively, this means that a player's hidden state gets affected less
by observed states that are very different from his/hers current hidden state.
For the experiments we show here we assume that $\Mmatrix(o = \ell, h =\ell)$ is the same for all values of $\ell\in\{-1,0,1\}$.  In this scenario any observed state of any player influences the others 
to stay in the corresponding hidden state. The particular matrix we used for our simulations in shown in Appendix~\ref{app:modelsettings}.

\spara{Setting $\Nmatrix$:} Matrix $\Nmatrix$ captures the influence of a player's hidden state at $t-1$ to their own hidden state at $t$. 
Our assumption is that each player is inclined to stay in the same hidden state; i.e., $\Nmatrix(h,h)> \Nmatrix(h,h')$, for all $h\neq h'$  and $h,h'\in\mathcal{H}$.

\spara{Setting $\Rmatrix$:} Parameters $\Rmatrix^i= (r_i\mid R_{i1}, \cdots, R_{in})$ for all $i\in [n]$
assign weights to the influence matrices $\Mmatrix$ and $\Nmatrix$ (see Eq.~\eqref{eq: cond-prob}).
We consider different scenarios for $\Rmatrix^i$:

\squishlist
\item $\Rmatrix_\text{H}$: All players depend on their previous hidden state. That is, $r_i = 1, \forall i \in [n]$ and $R_{ij}=0, \forall i,j\in [n]$. We denote the profile that corresponds to this $\Rmatrix$ with \texttt{H}.
\item $\Rmatrix_\text{HO}$: All players depend only on their previous hidden and observed states. That is, $r_i>R_{ii}>0,\forall i\in [n]$ and 
$R_{ij}=0, \forall i,j\in [n]$.
For our experiments we use $r_i = 0.7$ and $R_{ii} = 0.3$, for all $i \in [n]$. 

\item $\Rmatrix_k$: There are $k$ \emph{pillar players} $P$ that affect all others; {\ie}, $r_i=R_{ii}=0$  and $R_{ij}=1/|P|$ for
all $i\notin P$ and $j\in P$ and $r_i=1$ for $i\in P$.

\item $\Rmatrix_\text{kH}$: There are $k$ pillar players $P$ that affect all others, but all 
non-pillar players are also affected by their hidden states. That is, 
$r_i=1$ for all $i\in P$, $r_i>0$ for all $i\notin P$ and $R_{ij}=(1-r_i) 1/|P|$ for all $i\notin P$ and $j\in P$. In our experiments we set $r_i=0.5$ for all $i\notin P$.

\item $\Rmatrix_\text{kD}$: In this case, there are again $k$ pillar players $P$ (as in $\Rmatrix_k$, but this time they depend on each other. That is, for every $i\in P$ $0<r_i<1$
and $R_{ij}=(1-r_i)1/|P|$ for every $i,j\in P$. In our experiments we again set $r_i=0.5$.


\squishend

\spara{Setting $\Ematrix$:} Matrix $\Ematrix$ controls the emission probabilities.
We assume that the observed states are strongly correlated with hidden states. That is,
$\Ematrix(o=\ell, h = \ell)> \Ematrix(o=\ell, h=\ell')$ all $\ell,\ell'\in\{-1,0,1\}$ and $\ell\neq \ell'$.
For example, if a player is in a poor mental state, we expect their performance (observed state) to also be poor low.

Since we do not vary $\Mmatrix$, $\Nmatrix$ and $\Ematrix$, the instantiations of  $\Rmatrix$
give rise to different team profiles. Thus, we have the following mappings between the $\Rmatrix$ instances and team-profile names:
$\Rmatrix_\text{H}\rightarrow \texttt{H}$,
$\Rmatrix_\text{HO}\rightarrow \texttt{HO}$,
$\Rmatrix_k\rightarrow \texttt{kPillars} (\texttt{kP})$,
$\Rmatrix_\text{kH}\rightarrow \texttt{kPillars+H} (\texttt{kP + H})$ and 
$\Rmatrix_\text{kD}\rightarrow \texttt{kPillars+D} (\texttt{kP + D})$.

\subsection{Simulations}
In order to better understand our models we create 
data that correspond to $7$ distinct team profiles; for a larger set of team profiles
see Appendix~\ref{app:team_profiles}.
Using these profiles we generate $1000$ datasets for each profile, where 
each dataset has $n=5$, $T=100$ and every player $i\in [n]$ starts from hidden state $H_1^i = 0$. 

In Appendix~\ref{app:team_state} we show the  \emph{hidden} and \emph{observed team states} for the profiles we mention above and others we consider in the appendix. 
The summary of the results is the following: in almost all cases the hidden and the observed team states follow closely one another.
All teams' states converge to the expected performance (0). Some differences are observed between profiles with respect to the 25-75 percentiles of the values we get in the simulations.

\begin{figure}[H]
  \centering
\includegraphics[width=0.6\linewidth, height = 4cm]{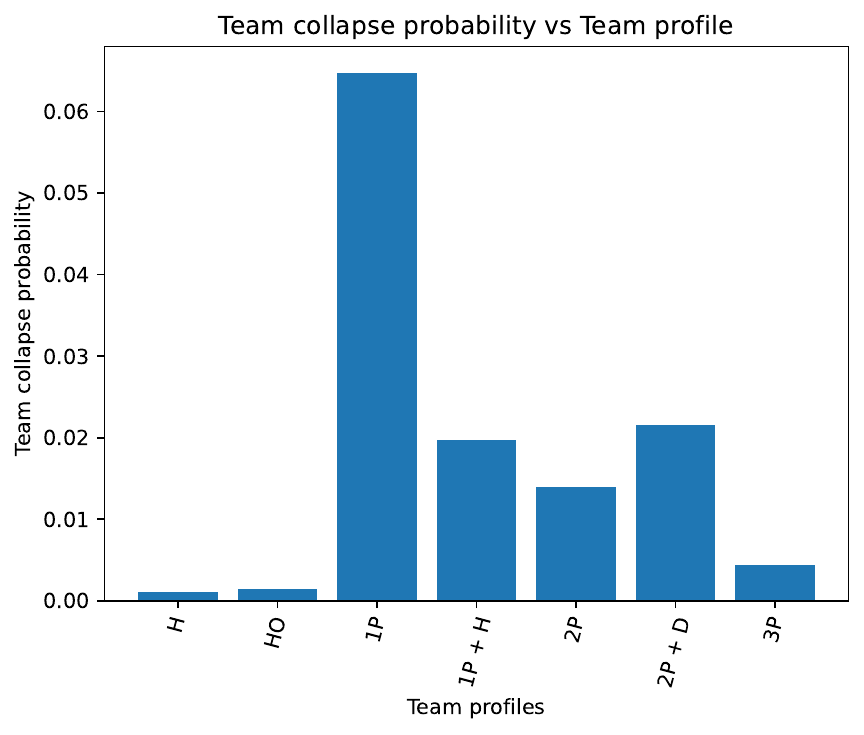}
  \caption{Team collapse probability vs team profile}
  \label{fig: team_collapse_prob}
\end{figure}

\spara{Probability of team collapse.}
In order to empirically estimate the probability of a team collapse for a given team, we count all the times the team collapsed in the simulations and divide by the total number of samples ($1000 \times 100 = 10^5)$. In these experiments, we consider
that $\alpha=1$ (see Dfn.~\ref{dfn:collapse}); {\ie} a team collapses if all its players have choked.
In Fig.~\ref{fig: team_collapse_prob} we show the 
team-collapse probability for the different team profiles we considered. 

First, we observe that when all players are completely unaffected by other players (profiles \texttt{H} and \texttt{HO}), then the team rarely collapses. In this case, the only way for a collapse to occur is all players independently under-performing at the same time.
In general, adding dependence on the hidden state of the players, {\ie}, making the players more self-dependent results in a more resilient team.

We also observe that the when there is one pillar in the team (profiles \texttt{1Pillar} and \texttt{1Pillar+H}), the team has a single point of failure.
That is, when the pillar happens to under-perform, the rest of the players are negatively affected and transition to their lowest mental (hidden) state.
Subsequently, these players also under-perform until the performance of the pillar player improves.  
Naturally, adding more pillars (profiles \texttt{kPillars}, ) makes the team more resilient to collapse since not all pillars under-perform at the same time.

Furthermore, dependence between pillar players ({\eg}, profile \texttt{2Pillars+D}) results in less resilient teams. This is because, dependent pillar players have a tendency to under-perform at the same time. Once this happens, the rest of the players transition to hidden state $-1$ and start under-performing.

Finally, we emphasize that a team that doesn't under-perform isn't necessarily a good team. Observed performance is measured relatively to the average performance of the team.
Thus, a consistently bad team might never under-perform, but still perform poorly relatively to other teams.


\subsection{Remarks on team profiles}
We summarize here our observations and state some principles we experimentally found to be true and can provide useful insights to  experts reasoning about team composition and resilience of teams to collapses. 

\begin{remark}
    (Self-dependence.) Self-dependent players, {\ie}, players that primarily depend on their own hidden and observed states, improve the resilience of teams.
\end{remark}

We say that a player is a \emph{pillar} if other players heavily depend on him/her, while he/she is self-dependent. For example, a player $i$ is a pillar player if $r_{i} \approx 1$ and $R_{ji} \approx 1, \forall j \in [n]$.

\begin{remark}
    (Pillars.) More pillars are better than less pillars.
\end{remark}

\begin{remark}
    (Dependence between pillars.) Dependence between pillars makes teams more vulnerable to team collapse.
\end{remark}

These observations have several implications in terms of team and roster construction. 
More specifically, self-dependent players are also typically players that can do several things at a high level. 
This is in contrast to ``specialists'' that bring to the team a very well-delineated skillset ({\eg}, individual defense and three point shooting) and depend on their teammates for other aspects of the game (e.g., shot generation). 
While lately there are many teams that prioritize the acquisition of specialists who possess specific skillset at an elite level, our observations indicate that having more self-dependent players makes a team more resilient to a possible collapse. 

Furthermore, as one might have expected, a team with self-dependent players that can also influence/help their teammates' performance ({\ie}, pillar players) is in general more resilient. 
However, the choice of these pillar players should take into consideration their dependence. 
According to our third remark, if a team includes pillars that are dependent, the resiliency to team collapse reduces. 
This dependency can be related to the position the two pillars cover on the team. 
For instance, if a team's pillar A is a center and its other pillar B is a guard there can be directional dependency from the B to A, just because B controls the ball more and is the one responsible for distributing it to other players. 
This dependency does not have to be strong, but its effects can trickle down to the rest of the players. 
Hence, from a roster-construction perspective, assuming a team already has one pillar, understanding the dependency that will be introduced when adding a second one is crucial. 
Going back to the example above, the best pillar pairing for player B might be another guard instead of a center.

%% file: sections/learning.tex
\section{Learning the model}
\label{sec: learning}
In this section, we describe a learning algorithm. 
Given the observed states of the players of a team over a period of time, it estimates
the parameters of our model and the corresponding hidden states of the players.

Throughout the description we follow the following notation:
we use $\Ob$ to denote the sequence of observations $\Ob_1, \cdots, \Ob_T$, where $\Ob_t = (O_t^1, \dots, O_t^n)$ are the observed signals of all entities at time $t$.
Accordingly, we use $\Hb$ to denote the sequence of hidden states $\Hb_1, \cdots, \Hb_t$,
where $\Hb_t = (H_t^1, \dots, H_t^n)$ are the hidden states of all entities at time $t$. Finally, we use $\Ob_{1:t}$ to denote $\Ob_1, \dots, \Ob_t$. 
Thus, given 
$\Ob$, we want to learn the parameters $\params=\{\Mmatrix^{ij}, \Nmatrix^i, \Rmatrix,\Ematrix^i\mid i,j\in [n] \}$ of the model and the sequence of hidden states $\Hb$.

\subsection{Learning the parameters of the model}
We learn the parameters of the model by applying the \emph{Expectation-Maximization} (EM) algorithm. EM alternates between two steps:  the expectation step (E-step) and the maximization step (M-step).
In our case, these steps are:

\textbf{E-step:} In this step, we need to calculate the \emph{expected complete log-likelihood}:
    \begin{align}\label{eq:q}
    Q(\params \mid \params^\text{old}) = 
    \EX_{\Hb \sim \Prob(\Hb \mid \Ob, \params^\text{old})}
    \left[
    \log \Prob(\Hb, \Ob \mid \params)
    \right],
\end{align}
    Usually, we approximate the above expectation by drawing samples from the \emph{posterior distribution} $\Prob(\Hb \mid \Ob, \params)$
    and calculating the average of the function.

\textbf{M-step}: In this step, we need to maximize the \emph{expected complete log-likelihood} w.r.t. the parameters of the model.

EM alternates between the E and the M steps until convergence. 

\subsubsection{E-step}
In the E-step we need to sample from the posterior distribution 
$\Prob(\Hb \mid \Ob, \params)$,
assuming that the model parameters $\params$ are known. Since it is clear from the context that
all probabilities will be conditioned on $\params$, we omit the  parameters in the 
condition and write the posterior as $\Prob(\Hb \mid \Ob)$

The key idea for being able to sample from the posterior is that we need to write
$\Prob(\Hb \mid \Ob)$ as a product where each 
random variable $\Hb_t$ is only conditioned on observed quantities or on random variables
$\Hb_{t'}$ with $t'<t$.  This is summarized in the following Lemma, which is the main result of this section:
\begin{lemma}\label{lemma:sampling}
    It holds that
    \begin{align}
        \Prob(\Hb \mid \Ob) = 
        \prod_{i=1}^n \prod_{t=1}^T 
        \frac{\Prob(H_t^i \mid \Ob_{1:t})}{\Prob(H_t^i \mid \Ob_{1:t-1})}
        \Prob(H_t^i \mid H_{t-1}^i, \Ob_{1:t-1})
    \end{align}
\end{lemma}
The proof of the lemma is based on a set of observations and it is quite lengthy, so it is given in Appendix~\ref{app:expectation}. This lemma allows us  to
 sample $H_t^i$ from a distribution that is proportional to
\[\frac{\Prob(H_t^i = h \mid \Ob_{1:t})
    \Prob(H_t^i = h\mid H_{t-1}^i, \Ob_{t-1})}{ \Prob(H_t^i = h \mid \Ob_{1:t-1})},\]
    as dictated by Lemma~\ref{lemma:sampling}.

\subsubsection{M-step}
In the M-step we want to maximize the 
the expected complete
log-likelihood w.r.t. the parameters of the model.
To see how this can be approached we first 
write the complete likelihood:
\begin{align*}
    \Prob(\Hb, \Ob) &= 
        \Prob(\Hb_1) 
        \prod_{t=2}^T 
        \Prob(\Hb_t \mid \Hb_{t-1}, \Ob_{t-1}) \prod_{t=1}^T 
        \Prob(\Ob_t \mid \Hb_t) \\
        &= 
        \Prob(\Hb_1) \prod_{t=2}^T \prod_{i=1}^n \Prob(H_t^i \mid H_{t-1}^i, \Ob_{t-1}) \prod_{t=1}^T 
        \prod_{i=1}^n \Prob(O_t^i \mid H_t^i)
\end{align*}
Taking the logarithm on both sides we get the complete log-likelihood:
\begin{align}\label{eq:completeloglikelihood}
    \log \Prob(\Hb, \Ob) &=
    \log \Prob(\Hb_1) 
    \\
    &+\sum_{t=2}^T \sum_{i=1}^n \log \Prob(H_t^i \mid H_{t-1}^i , \Ob_{t-1}) \nonumber \\
    &+ 
    \sum_{t=1}^T \sum_{i=1}^n \log \Prob(O_t^i \mid H_t^i).\nonumber
\end{align}
Thus in the
M-step we have to solve the following problem:
\begin{align}
\params^\text{new} = \argmax_{\params} Q(\params \mid \params^\text{old}),
\end{align}
where $Q()$ is given by Eq.~\eqref{eq:q} and the complete log-likelihood is computed by
Eq.~\ref{eq:completeloglikelihood}.
Note that $ Q(\params \mid \params^\text{old})$ is not a concave function in the parameters $\params$ and thus finding a global maximum it not easy.
Specifically, the problem stems from the terms $\Prob(H_t^i \mid H_{t-1}^i , \Ob_{t-1})$ which contain products of variables. Thus, we opted for a round-robin heuristic algorithm. This heuristic
iteratively optimizes one parameter 
from $\params=\{\Mmatrix, \Nmatrix, \Rmatrix^i,\Ematrix \mid i\in [n] \}$
per iteration, 
while keeping the other parameters fixed;
in all cases fixing all but one of the parameters results in a concave function. Hence, we can use standard convex-optimization tools to optimize for each parameter separately. In practice, we found that 1-2 iterations of the round robin heuristic are adequate.

\subsection{Likelihood calculation}
Given the model parameters $\params$, we can calculate the likelihood of the data $\Pr(\Ob)$, based on the results of the previous section:
\begin{align}
    \log \Pr(\Ob) &= 
        \sum_{t=1}^T \sum_{i=1}^n
            \log \Pr(O_t^i \mid \Ob_{t-1}) \\
        &= \sum_{t=1}^T \sum_{i=1}^n \log \left[ \sum_{h \in \mathcal{H}} \Pr(O_t^i \mid H_t^i = h) 
        \Pr(H_t^i = h \mid \Ob_{1:t-1}) \right].
\end{align}

\subsection{Decoding the hidden states}\label{sec:decoding}
In decoding , we find the 
most likely hidden variables path $\Hb_1, \dots, \Hb_T$, given 
the model parameters $\params$ (evaluated by EM) and the observed variables $\Ob = \Ob_{1:T}$. Formally, this problem can be expressed as 
\begin{align}\label{eq:decoding}
    \overline{\h} = \argmax_{\h} \Prob(\Hb = \h \mid \Ob).
\end{align}
We solve the problem expressed in Eq.~\eqref{eq:decoding} via a dynamic-programming 
algorithm, which is very similar to the \emph{Viterbi} algorithm used for standard HMMs~\cite{bishop06pattern}.
To this end, first we decouple the chains
\begin{align*}    
    \Prob(\Hb = \h \mid \Ob) =
    \prod_{i=1}^n \Prob(\Hb^i = \h^i \mid \Ob),
\end{align*}
using the independence of $\Hb^i$ and $\Hb^j$ given $\Ob$.
For each entity $i$, we define $\delta_i(t,h)$ to be the probability of the 
maximum-probability path ending at $H_t^i=h$; formally:
\begin{eqnarray*}
\lefteqn{
    \delta_i(t,h) =  \max_{h_1, \dots, h_{t-1} \in \mathcal{H}^{t-1}}}\\ && \Prob (H_1^i = h_1, \cdots, H_{t-1}^i = h_{t-1}, H_t^i = h \mid O_1^i, \cdots O_t^i).
\end{eqnarray*}
That is, $\delta_t(h)$ is the probability of the most probable path ending at $H_t^i = h$.
Now the values of $\delta_i(t,h)$ can be computed using the following dynamic-programming recursion:
\begin{eqnarray*}
    \lefteqn{ \delta_i(t,h) = \Prob(O_t^i \mid H_t^i = h) \times}\\ && \max_{h^\prime \in \mathcal{H}} \left( \delta_i(t-1,h^\prime) \Prob (H_t^i = h \mid H_{t-1}^i = h^\prime, \Ob_{t-1}) \right),
\end{eqnarray*}
$\forall h \in \mathcal{H}$ and $t\in \{1,\ldots , T\}$.

The base case is
\begin{align*}
    \delta_i(1,h) = \Prob(H_1^i = h) \Prob(O_1^i \mid H_1^i = h), \forall h \in \mathcal{H}.
\end{align*}

Once all $\delta_i(t,h)$ values are computed one can trace back the solutions
starting from $\delta_i(T,h)$, where $h=\argmax_{h^\prime \in \mathcal{H}}\delta_i(T,h^\prime)$.


\subsection{Evaluating the learning algorithm on synthetic data}
\label{sec: EM synthetic}
\begin{figure}[h]
  \centering
\includegraphics[width=0.6\linewidth, height = 4cm]{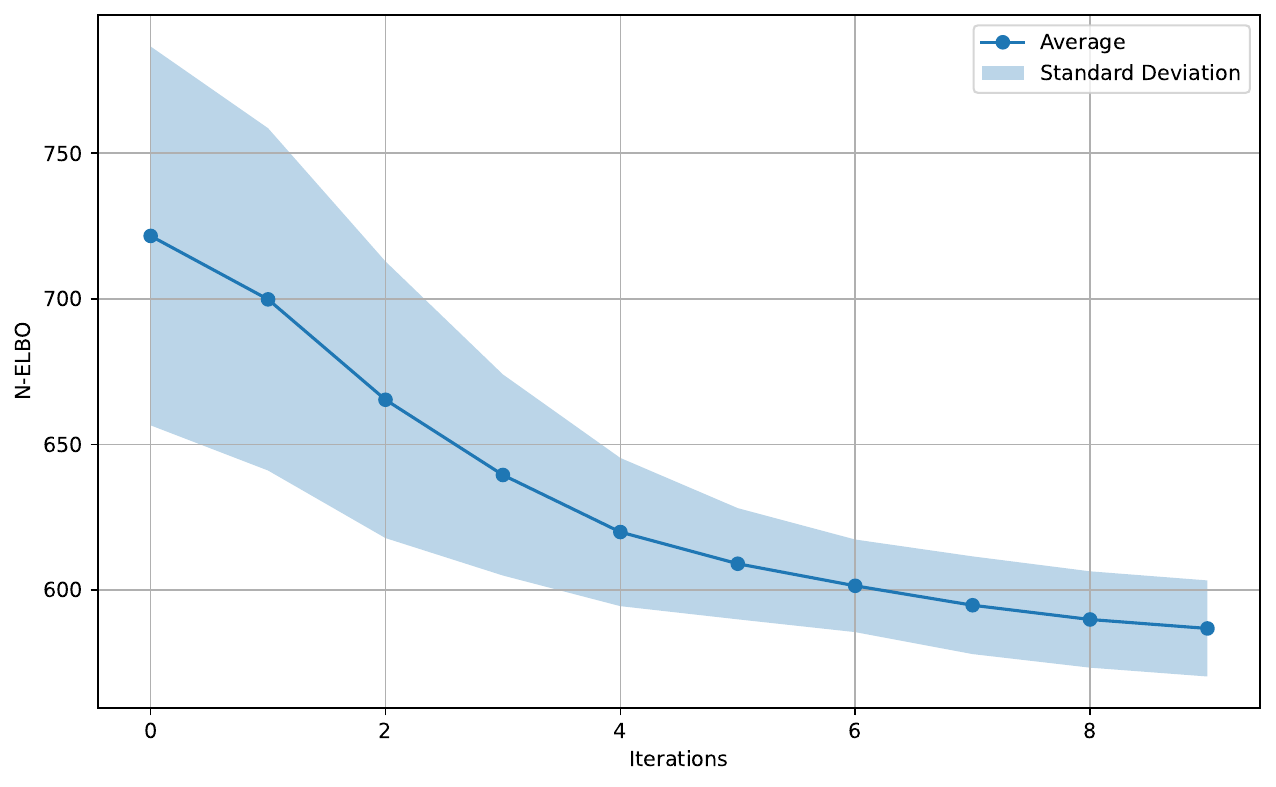}
  \caption{Negative Evidence Lower Bound (N-ELBO) vs iterations of EM; average and standard deviation over 10 random initializations.}
  \Description{Negative Evidence Lower Bound (N-ELBO) vs iterations of EM}
  \label{fig: ELBO synthetic}
\end{figure}

We evaluated our implementation of the EM algorithm on synthetic data as follows.
We created a \texttt{1Pillar} team profile and generated 
samples that correspond to $T=100$ timesteps. We then ran the EM algorithm and found that it correctly identifies the structure of the team.
Figure~\ref{fig: ELBO synthetic} shows the Negative Evidence Lower Bound (N-ELBO) as a function
of the iterations of EM. 
The figure shows the average and the standard deviation of N-ELBO over $10$ different initializations.
As expected, 
N-ELBO decreases with iterations until 
it stabilizes to a low value.
The negative likelihood of the (best) learned model is $416$, while the negative likelihood of the true model is $417$ (slightly larger). That is, EM finds a model 
that has even better likelihood than the model used to generate the data.

%% file: sections/experiments.tex
\begin{figure*}
    \centering
    \begin{subfigure}[b]{\figwidth}
        \centering
     \includegraphics[width=0.8\textwidth]{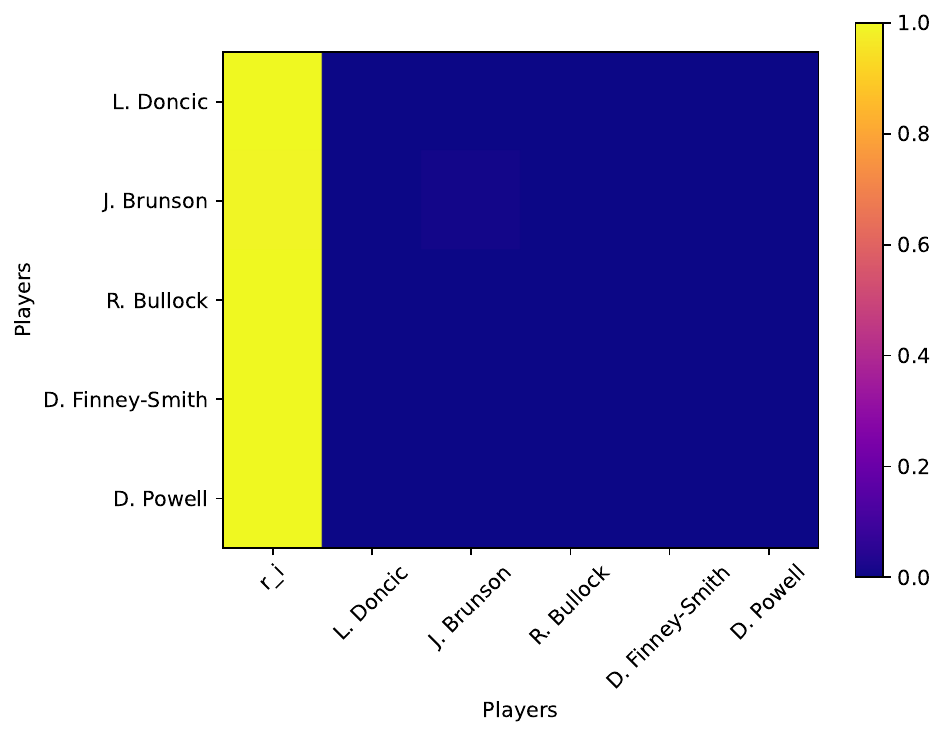}
        \caption{Dallas Mavericks}
        \label{fig: NBA team profiles main - mavs}
    \end{subfigure}
     \begin{subfigure}[b]{\figwidth}
        \centering
     \includegraphics[width=0.8\textwidth]{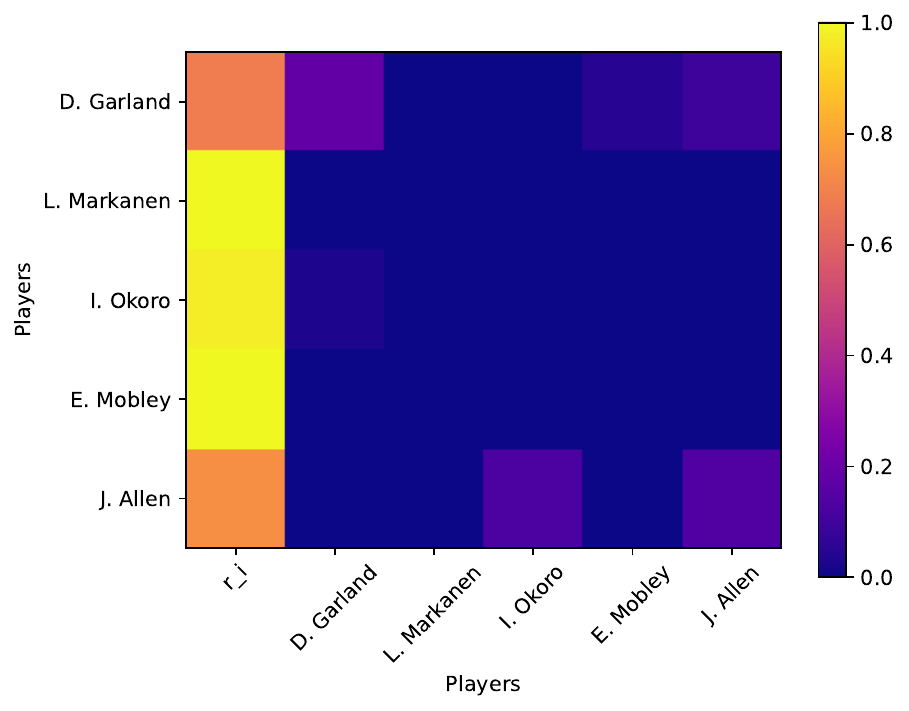}
        \caption{Cleveland Cavaliers}
        \label{fig: NBA team profiles main - cavs} 
    \end{subfigure}
    \begin{subfigure}[b]{\figwidth}
        \centering
    \includegraphics[width=0.8\textwidth]{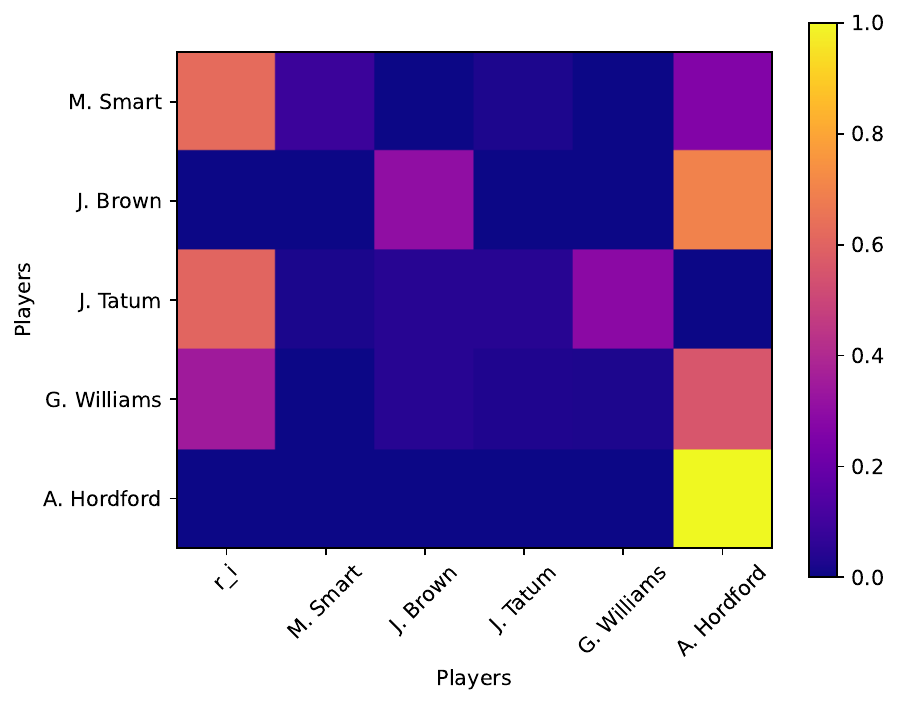}
        \caption{Boston Celtics}
    \label{fig: NBA team profiles main - celtics}
    \end{subfigure}
     \begin{subfigure}[b]{\figwidth}
        \centering
     \includegraphics[width=0.8\textwidth]{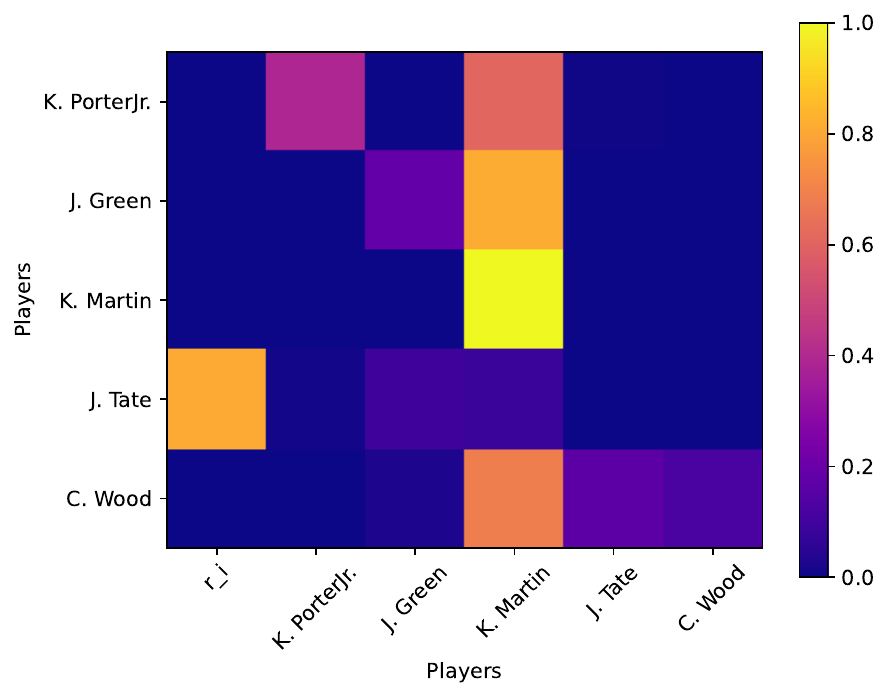}
        \caption{Houston Rockets}
        \label{fig: NBA team profiles main - rockets}
    \end{subfigure}
    \caption{Team structure ($\Rmatrix$) for the Dallas Mavericks, Cleveland Cavaliers, Boston Celtics and Houston Rockets.}
    \label{fig: NBA team profiles main}
\end{figure*}

\section{Real-world experiments}
\label{sec: real-world experiments}
In this section, 
we validate the practical utility of our model by
using it to identify the structure of all NBA teams for the season 2021-2022. We also show how our model can be used to analyze and obtain significant insights for specific games.

\subsection{Data}
\label{sec: NBA Data}
For each NBA team of the 2021-2022 season we selected their  starting lineup players. That is, the players that played the most minutes in each position (Point Guard, Shooting Guard, Small Forward, Power Forward, Center)\footnote{The starting lineups are obtained from the basketball reference website. For example, for Boston Celtics the starting lineup is here:  https://www.basketball-reference.com/teams/BOS/2022\_depth.html}.

For our experiment,s we used a Play-by-Play dataset (\PBP) that contains a detailed record of events for all NBA games of the season. Using {\PBP},
for each player, we computed his average {\PIE} metric\footnote{See https://www.nba.com/stats/help/faq} across all quarters (each game has four 12min quarters) of all games. At a high level, the {\PIE} metric of a player for a given time window, captures the percentage of events (points, assists, rebounds, etc) that the player achieved in the given window.
Finally, we labeled the performance of every player for every quarter he played as average ($O_t^i = 0$), if his {\PIE} metric was within $\pm \epsilon$ of his average {\PIE}, where $\epsilon = 0.05$. If his performance was above this region, we labeled his performance as over-performance $O_t^i = 1$. Accordingly, if his performance was below the average region we labels his performance as under-performance $O_t^i = -1$.
If the player didn't play in a quarter we used his latest {\PIE} measurement in the game, or if there was no previous measurement we assumed average performance $O_t^i = 1$. We name the resulting dataset \NBAseason.

For section \ref{sec: diving deeper}, we created the {\NBAgames} dataset. This dataset contains a few selected games of interest. We divided each of these games into 3 minute time windows so as to have more fine-grained measurements. As before, for each player we computed his average {\PIE} metric in each 3 minute window and labeled his performance.

\subsection{Implementation details}
\label{sec: implementationdetails}
In order to apply EM in real data, we created a collection of approximately $10000$ team profiles, \ie, instantiations of our model. 
These profiles were, to a large extent, informed by our findings in Section~\ref{sec: EM synthetic}. 
Then, we found the profile from the collection that best matches each team by calculating the likelihood of the profile.
Finally, for each team we ran the EM algorithm initialized with the aforementioned profile.


The implementation of our algorithms, along with the datasets we used in our experiments are publicly available at \url{https://github.com/jasonNikolaou/Team-collapse.git}.

\subsection{Team structure in NBA teams}
\label{sec: team structure}
Using dataset {\NBAseason}, we applied the EM algorithm to find the maximum-likelihood estimates of the parameters of our model as described in Section~\ref{sec: implementationdetails}.
\begin{figure*}[t]
    \centering
    \begin{subfigure}[b]{0.33\textwidth}
        \centering
     \includegraphics[width=\textwidth]{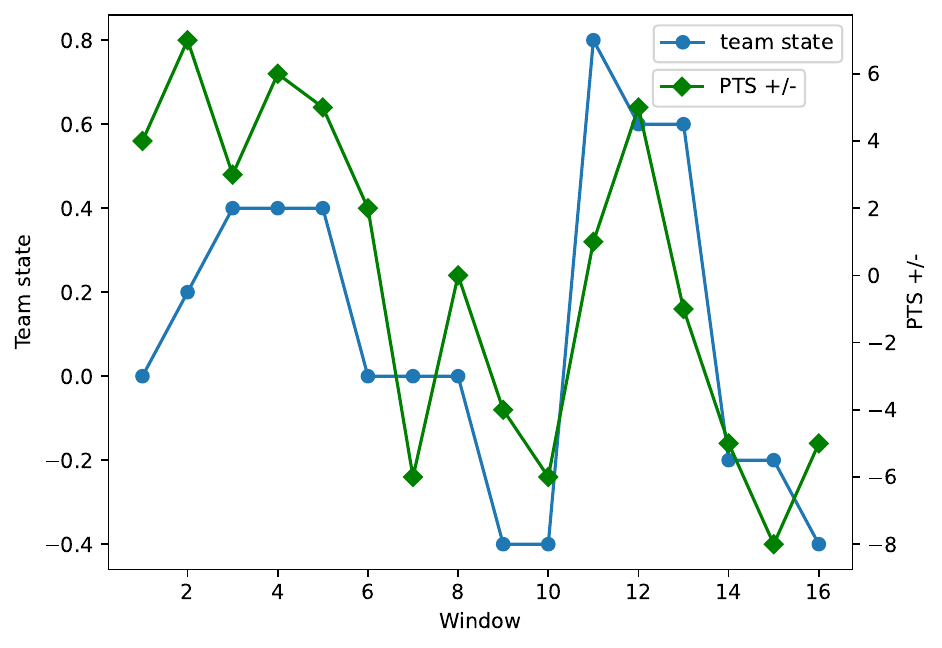}
        \caption{Nets vs Knicks game (16 Feb 2022)}
         \label{fig: BKN_NYK game}
    \end{subfigure}
    \hfill
    \begin{subfigure}[b]{0.33\textwidth}
        \centering
    \includegraphics[width=\textwidth]{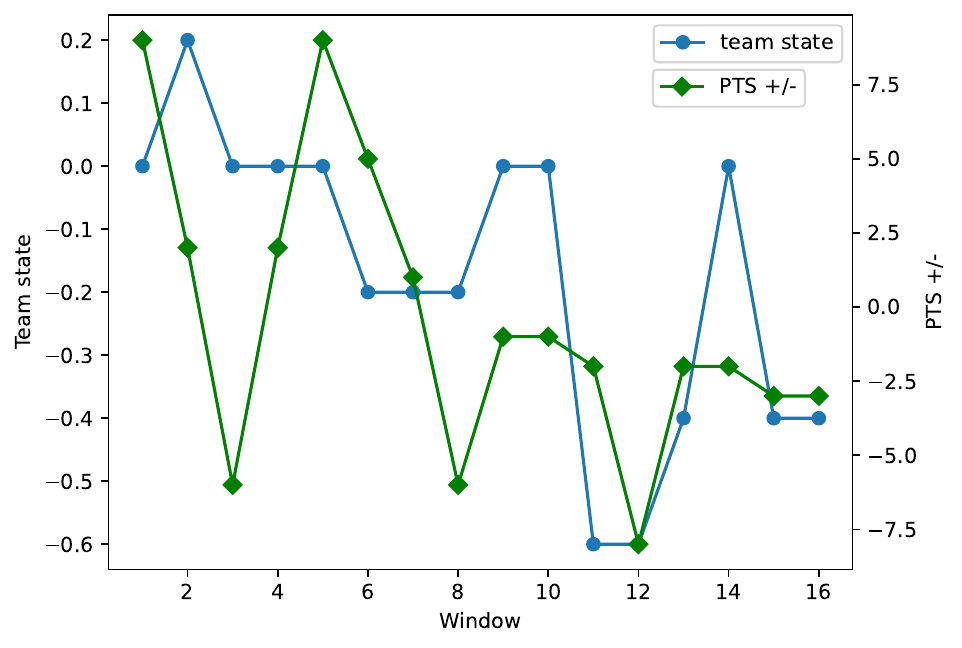}
        \caption{Celtics vs Knicks game (6 Jan 2022)}
        \label{fig: BOS_NYK game}
    \end{subfigure}
        \hfill
    \begin{subfigure}[b]{0.33\textwidth}
        \centering
    \includegraphics[width=\textwidth]{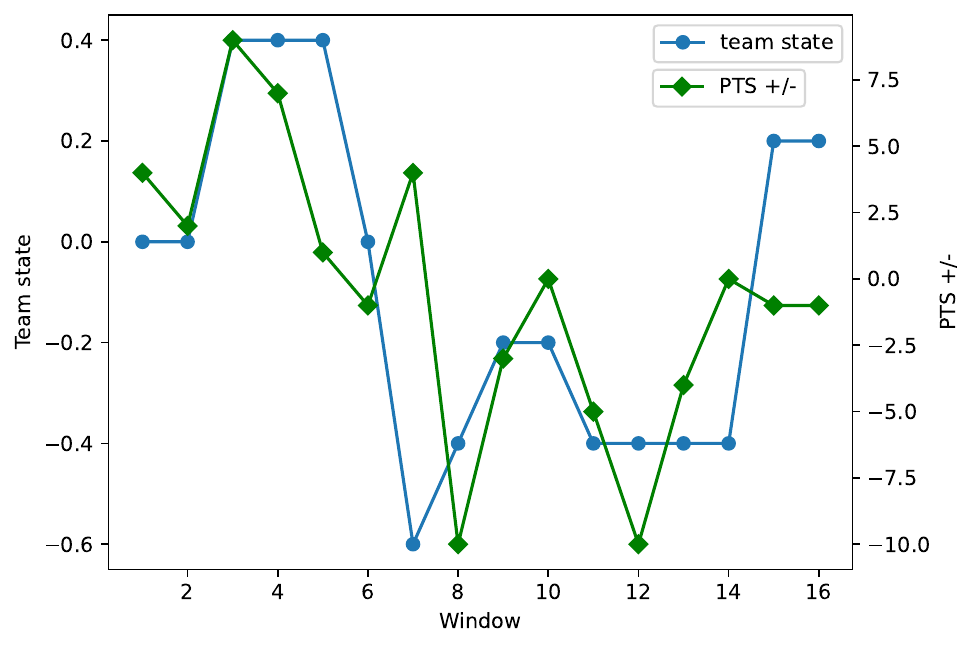}
        \caption{Lakers vs Thunder (27 Oct 2021)}
        \label{fig: LAL_OKC game}
   \end{subfigure}
    \caption{Team (hidden) state and points difference (PTS +/-) per 3 minute window for different games.}
    \label{fig:example_games}
\end{figure*}

Thus, for every one of the 30 NBA teams we obtained an estimate
of its \emph{team structure} $\Rmatrix$.
While of course each team has a different 
structure, 
there are two main team-structure types that emerge.   
In Figure \ref{fig: NBA team profiles main} we present four representative teams of these  team structures (two for each), while we present the rest of the team structures in Appendix \ref{app: team-profiles}. 

The team structures correspond to the matrix $\Rmatrix$ of the team profile 
and therefore, they 
are presented as a $5\times 6$ heatmap. The first column in the heatmap corresponds to the dependency of each row player $i$ to its hidden state ({\ie}, $r_i$), and the rest $5\times 5$ matrix represents the dependency between the hidden state of the row player $i$ 
and the observed state of the column player $j$.

The first two team structures, 
shown in Figures~\ref{fig: NBA team profiles main - mavs}
and~\ref{fig: NBA team profiles main - cavs} correspond to 
the structures of Dallas Mavericks and Cleveland Cavaliers respectively.
For the Mavericks, we see that the state of every player is mainly dependent on their own hidden state (first column of the matrix), while they do not depend on their own observations. 
Simply put, these teams consist of players that typically can showcase ``short memory'' with respect to their observed performance/state and their state only depends on their own hidden (mental) state.
The Cavaliers' team structure is very similar.

The other two team structures shown in Figures~\ref{fig: NBA team profiles main - celtics} and~\ref{fig: NBA team profiles main - rockets} correspond to Boston Celtics and Houston Rockets respectively.
For Celtics, we see a smaller dependency of their players' on their own hidden state and a larger dependency for some of them on their observed state. 
However, there is also a dependency on another player's observed state (last column in Figure \ref{fig: NBA team profiles main - celtics}). 
In this particular case , this player is Al Horford, who is the most experienced player on the team and his teammates can be thought of as ``feeding off'' his observed state/performance. 
The Rockets team structure (Figure~\ref{fig: NBA team profiles main - rockets}) is very similar.

Interestingly, while both the Dallas Mavericks and Boston Celtics have very different structures, they both reached the conference finals and finals respectively. 
On the other hand, the Houston Rockets, which have structure
similar to  Boston Celtics, had a record of 20 wins and 62 loses and finished last in the Western conference. 
This demonstrates that there are many ways to succeed (or fail) in a team sport. The team structure is just one characteristic of a team, among others that create different vulnerabilities and strengths.

\subsection{Diving deeper into specific games}
\label{sec: diving deeper}
For the results we report here, we use the {\NBAgames} dataset described in section \ref{sec: NBA Data}.
Using the parameters learned in the previous section, we decode the most probable hidden states of players by applying the decoding algorithm described in Section~\ref{sec:decoding}.

We now analyze 
three games in detail. In all these games, one of the teams managed to comeback and win the game despite trailing behind in score for the most part of the game. More analyzed games can be found in the Appendix \ref{app: NBA games}. 

Figure \ref{fig:example_games} visualizes the hidden team state we obtained from our model over the 3 minute windows, and overlays the actual point differential for the game during these states. 
In these games, a team was performing well to start the game (Knicks, Celtics and Lakers respectively in the figure order), but later in the game there was a ``transition'' and they surrendered the lead and ultimately lost the game (i.e., collapsed). 
As we can see the inferences on the team state made by our model follow closely the point differential, and hence the status, of the game. 
These results indicate that our model captures the underlying hidden states of the players to a great extent and can explain what one observes at the game.

%% file: sections/conclusions.tex
\section{Conclusions}
In this paper, we proposed a probabilistic graphical model that allowed us to 
formalize the dependencies between team players' 
hidden/mental states and the observed performance of themselves and their teammates. 
Then, we used this model
to identify, analyze and explain team collapses, i.e., events where a team severely under-performed.
We also showed that we can learn the parameters of this model using an EM algorithm. Furthermore, we used our model to analyze real-world team data from NBA games in 2021-22 season. As a result, we found some interesting team structures and identified games where according to our model a team collapsed.

One shortcoming of our model is the assumption that players in the team do not change over time. Considering a dynamic version of the model, where players can get substituted is an interesting direction of future work.
In our experiments with real-world data, we relied on the use 
of the PIE metric to measure the observed performance of players. In the future, we plan to explore other metrics as well as make use of machine learning methods in order to classify the observed performance.
Finally, we can potentially use our model to calculate collapse probabilities, as predictors of future collapses. In such a scenario, our model can guide coaching decisions, {\eg}, substitutions, time-out, etc.

\textbf{Reproducibility.} We make the code and the data we used in our experiments publicly available at \url{https://github.com/jasonNikolaou/Team-collapse.git}.

%% file: sections/appendix.tex
\section{Appendix for Section~\ref{sec:model}}~\label{app:model}

\subsection{Proof of Lemma \ref{lemma:model_probability_law}}~\label{app:problaw}
\begin{proof}
    \begin{align*}
        &\sum_{h \in \mathcal{H}} (H_t^i = h \mid \Ob_{t-1}, H_{t-1}^i) \\
        &=\sum_{h \in \mathcal{H}} \left( 
        \sum_{j=1}^n 
        R_{ij} \I(H_t^i = h \mid O_{t-1}^j)
        + R_i \I(H_t^i = h \mid H_{t-1}^i)
        \right) \\
         &=\sum_{j=1}^n R_{ij} \sum_{h \in \mathcal{H}} \I(H_t^i = h \mid O_{t-1}^j)
        + R_i \sum_{h \in \mathcal{H}} \I(H_t^i = h \mid H_{t-1}^i)\\
        &= \sum_{j=1}^n R_{ij} + R_i\\
        &= 1.
    \end{align*}
\end{proof}

\subsection{On the influence matrices} \label{app:influence_matrices}
In this section, we answer the following question: why isn't $\Mmatrix$ the same as the probability distributions $\Prob(H_t^i \mid O_{t-1}^i)$.
Consider the following proof by contradiction. Let $n=1$ and $r_1 = 0$, $R_{11} = 1$.
We define the following parameters:
\begin{align}
    \Nmatrix = 
    \begin{bmatrix}
        0 & 0 & 1 \\
        0 & 1 & 0 \\
        1 & 0 & 0
    \end{bmatrix},
    \Nmatrix = 
    \begin{bmatrix}
        1 & 0 & 0 \\
        0 & 1 & 0 \\
        0 & 0 & 1
    \end{bmatrix},
    \Ematrix = 
    \begin{bmatrix}
        0.7 & 0.2 & 0.1 \\
        0.2 & 0.6 & 0.2 \\
        0.1 & 0.2 & 0.7
    \end{bmatrix},
\end{align}
Assume that:
\begin{align}
    \Prob(H_t^i = h \mid O_{t-1}^i = o) = \Mmatrix[h, o]
\end{align}
We have
\begin{align}
    \Prob(H_t^1 = 0 \mid H_{t-1}^1 = 1, O_{t-1}^1 = 2)
        &= r_1 \Nmatrix[1, 0] + R_{11} \Mmatrix[2, 0] \\
        &= 0 + 1 \times 1 \\
        &= 1
\end{align}
and
\begin{align}
    \Prob(H_t^1 = 0 \mid H_{t-1}^1 = 1) = \Nmatrix[1, 0] = 0.
\end{align}
We also have
\begin{align}
    \Prob(H_t^1 = 0 \mid H_{t-1}^1 = 1) &=
        \sum_{o} \Prob(H_t^1 = 0 \mid H_{t-1}^1 = 1, O_{t-1}^1 = o) \Prob(O_{t-1}^1 = o \mid H_{t-1}^1 = 1) \\
        &\geq \Prob(H_t^1 = 0 \mid H_{t-1}^1 = 1, O_{t-1}^1 = 2) \Prob(O_{t-1}^1 = 2 \mid H_{t-1}^1 = 1) \\
        &\geq 1 \times 0.2 \\
        &= 0.2,
\end{align}
which is a contradiction.

In the same way, we can prove that $\Nmatrix$ isn't the same as the probability distributions $\Prob(H_t^i \mid H_{t-1}^i)$.

\section{Appendix for Section~\ref{sec:expressivity}}\label{app:expressivity}

\subsection{Examples of model parameters  used in our experiments}\label{app:modelsettings}

\spara{{\Nmatrix} instance used in the experiments presented in Section~\ref{sec:expressivity}:}
\begin{align}
    \Nmatrix = 
    \begin{bmatrix}
        0.7 & 0.2 & 0.1 \\
        0.2 & 0.6 & 0.2 \\
        0.1 & 0.2 & 0.7
    \end{bmatrix}
\end{align}

\spara{{\Ematrix} instance used in our experiments:}
\begin{align}
    \Ematrix = 
    \begin{bmatrix}
        0.8 & 0.2 & 0 \\
        0.2 & 0.8 & 0.2 \\
        0 & 0.2 & 0.8
    \end{bmatrix}
\end{align}

\spara{{\Mmatrix} instances used in our experiments in Section~\ref{sec:expressivity}}
\begin{align}
    \Mmatrix = 
    \begin{bmatrix}
        0.8 & 0.2 & 0 \\
        0.2 & 0.8 & 0.2 \\
        0 & 0.2 & 0.8
    \end{bmatrix},
\end{align}

\spara{Additional {\Mmatrix} instances we considered:}

\begin{align}
    \Mmatrix = 
    \begin{bmatrix}
        0.9 & 0.1 & 0 \\
        0.2 & 0.6 & 0.2 \\
        0.2 & 0.3 & 0.5
    \end{bmatrix}
\end{align}

and

\begin{align}
    \Mmatrix_{\text{anti}} = 
    \begin{bmatrix}
        0 & 0.2 & 0.8 \\
        0.1 & 0.8 & 0.1 \\
        0.8 & 0.2 & 0
    \end{bmatrix}.
\end{align}

\spara{Additional $\Rmatrix$ instances we considered}
In addition to the $\Rmatrix$ instances we described in Section~\ref{sec:expressivity}, we also experimented with a few more
$\Rmatrix$ instances.  We describe the full set of those instances here:

\squishlist
\item $\Rmatrix_\text{H}$: All players depend on their previous hidden state. That is, $r_i = 1, \forall i \in [n]$ and $R_{ij}=0, \forall i,j\in [n]$. We denote the profile that corresponds to this $\Rmatrix$ with \texttt{H}.
\item $\Rmatrix_\text{HO}$: All players depend only on their previous hidden and observed states. That is, $r_i>R_{ii}>0,\forall i\in [n]$ and 
$R_{ij}=0, \forall i,j\in [n]$.
For our experiments we use $r_i = 0.7$ and $R_{ii} = 0.3$, for all $i \in [n]$. 

\item $\Rmatrix_k$: There are $k$ \emph{pillar players} $P$ that affect all others; {\ie}, $r_i=R_{ii}=0$  and $R_{ij}=1/|P|$ for
all $i\notin P$ and $j\in P$ and $r_i=1$ for $i\in P$.

\item $\Rmatrix_\text{kH}$: There are $k$ pillar players $P$ that affect all others, but all 
non-pillar players are also affected by their hidden states. That is, 
$r_i=1$ for all $i\in P$, $r_i>0$ for all $i\notin P$ and $R_{ij}=(1-r_i) 1/|P|$ for all $i\notin P$ and $j\in P$. In our experiments we set $r_i=0.5$ for all $i\notin P$.

\item $\Rmatrix_\text{kD}$: In this case, there are again $k$ pillar players $P$ (as in $\Rmatrix_k$, but this time they depend on each other. That is, for every $i\in P$ $0<r_i<1$
and $R_{ij}=(1-r_i)1/|P|$ for every $i,j\in P$. In our experiments we again set $r_i=0.5$.

\item $\Rmatrix_\text{kDH}$: This is a combination of $\Rmatrix_H$ and $\Rmatrix_{kD}$ where all 
pillar players are set as in $\Rmatrix_{kD}$ and all non-pillar players are set as in $\Rmatrix_\text{kH}$.

\item $\Rmatrix_\text{uniform}$: In this case we assume that for every player $i$, $r_i=R_{ij}=1/(n+1)$.
\squishend

\subsection{Team profiles}\label{app:team_profiles}
We create the following extended sets of profiles:

\texttt{H}: {\Mmatrix} as above and $\Rmatrix^i=\Rmatrix_H$.

\texttt{HO.} {\Mmatrix} as above and $\Rmatrix^i=\Rmatrix_\text{HO}$. 

\texttt{1Pillar (1P).} {\Mmatrix} as above and $\Rmatrix^i=\Rmatrix_\text{1}$.

\texttt{1Pillar + Hidden (1P + H).} 
{\Mmatrix} as above and $\Rmatrix^i=\Rmatrix_\text{1H}$.

\texttt{2Pillars (2P).} 
{\Mmatrix} as above and $\Rmatrix^i=\Rmatrix_\text{2}$.

\texttt{2Pillars + Dependence (2P + D).} 
{\Mmatrix} as above and $\Rmatrix^i=\Rmatrix_\text{2D}$.

\texttt{2Pillars + Hidden (2P + H).} 
{\Mmatrix} as above and $\Rmatrix^i=\Rmatrix_\text{2H}$.

\texttt{2Pillars + Hidden + Dependence (2P + H + D).} 
{\Mmatrix} as above and $\Rmatrix^i=\Rmatrix_\text{2DH}$.

\texttt{3Pillars (3P).} 
{\Mmatrix} as above and $\Rmatrix^i=\Rmatrix_\text{3}$.

\texttt{3Pillars + Dependence (3P + D).} {\Mmatrix} as above and $\Rmatrix^i=\Rmatrix_\text{3D}$.

\texttt{3Pillars + Hidden (3P + H).} {\Mmatrix} as above and $\Rmatrix^i=\Rmatrix_\text{2H}$.

\texttt{3Pillars + Hidden + Dependence (3P + H + D).}
{\Mmatrix} as above and $\Rmatrix^i=\Rmatrix_\text{2DH}$.

\texttt{Uniform.} Players equally depend on their hidden states and other players, i.e. $R_i = 1/6, R_{ij} = 1/6, \forall i, j \in [n]$.

\texttt{1Pillar Bad Teammate (BT1P).} Same as the \texttt{1Pillar} profile, but we change the $\Mmatrix$ matrix to the following:
\begin{align}
    \Mmatrix = 
    \begin{bmatrix}
        0.9 & 0.1 & 0 \\
        0.2 & 0.6 & 0.2 \\
        0.2 & 0.3 & 0.5
    \end{bmatrix}.
\end{align}

We are trying to capture a pillar who has significantly negative impact when his/her observed state is $-1$ (i.e. below-average performance), and weakly positive impact when their observed state is $1$ (i.e. above-average performance).

\texttt{1Pillar Bad Teammate + Hidden (BT1P).}
Same as the \texttt{1Pillar + Hidden} profile, but with the $\Mmatrix$ matrix of the \texttt{BT1P} profile.

\texttt{1Pillar Great Teammate (GT1S).}
Same as the \texttt{1Pillar} profile, but we change the $\vect{M}$ matrix to the following:
\begin{align}
    \Mmatrix = 
    \begin{bmatrix}
        0.5 & 0.3 & 0.2 \\
        0.1 & 0.8 & 0.1 \\
        0 & 0.1 & 0.9
    \end{bmatrix}.
\end{align}

This profile captures a pillar who has significantly positive impact when their observed state is $1$ (i.e. above-average performance), and weakly negative impact when their observed state is $-1$ (i.e. below-average performance).

\texttt{1Pillar Great Teammate + Hidden (GT1P).} Same as the 
\texttt{1Pillar + Hidden} profile, but with the $\Mmatrix$ matrix of the \texttt{1-Pillar Great Teammate} profile.

Finally, we create two profiles to capture anti-dependence. That is, when one player's performance improves, another player's performance drops.
We define the anti-dependence $\Mmatrix$ matrix as:
\begin{align}
    \Mmatrix^{\text{antiD}} = 
    \begin{bmatrix}
        0 & 0.2 & 0.8 \\
        0.1 & 0.8 & 0.1 \\
        0.8 & 0.2 & 0
    \end{bmatrix}.
\end{align}

\texttt{1Pillar + anti-Dependence (1Pillar + antiD).} 
Same as the \texttt{1Pillar} profile, but we change the $\Mmatrix$ to $\Mmatrix^{\text{antiD}}$.

\texttt{2Pillars + anti-Dependence (2Pillars + antiD).}
Same as the \texttt{2Pillar} profile, but we change the $\Mmatrix$ matrix of both pillars  to $\Mmatrix^{\text{antiD}}$. In this way, the two pillars are anti-dependent on each other. Note that the pillars don't depend at all on non-pillar players.

\subsection{Team state over time} \label{app:team_state}
In order to better understand our models we experiment with the 19 profiles described in Appendix~\ref{app:team_profiles}. 
For each profile we types of teams and then run $1000$ simulations. Each simulation consists of $100$ samples.
We assume that every player $i\in [n]$ starts from hidden state $H_1^i = 0$. In our experiments,
we use 19 different team profiles, which correspond to different combinations of $\Rmatrix$ and
$\Mmatrix$ pairs. 

\begin{figure*}[htbp]
    \centering
    \begin{subfigure}[b]{\figwidth}
        \centering
     \includegraphics[width=\textwidth]{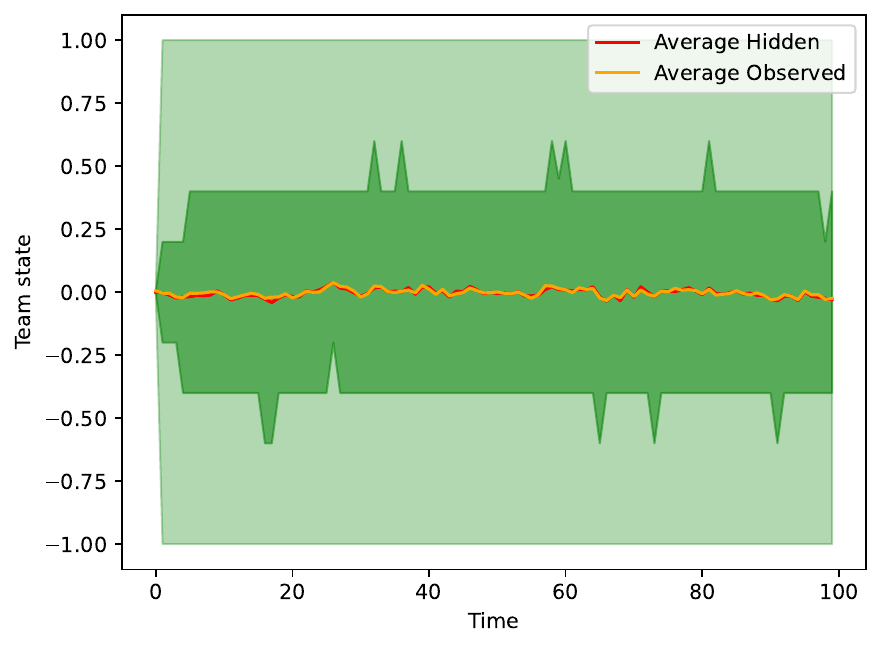}
        \caption{1Pillar}
    \end{subfigure}
    \hfill
    \begin{subfigure}[b]{\figwidth}
        \centering
    \includegraphics[width=\textwidth]{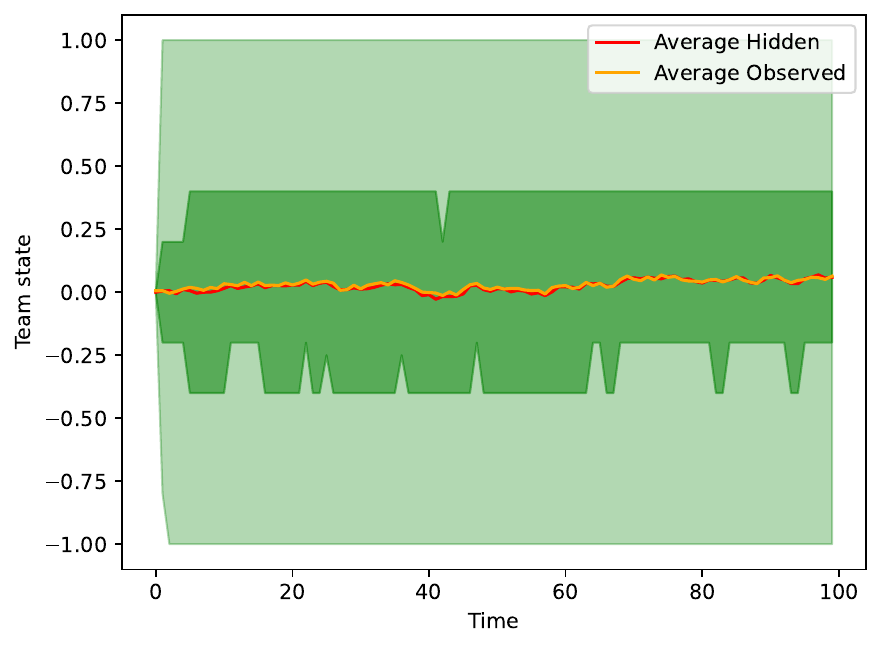}
        \caption{1Pillar + H}
    \end{subfigure}

    \begin{subfigure}[b]{\figwidth}
        \centering
     \includegraphics[width=\textwidth]{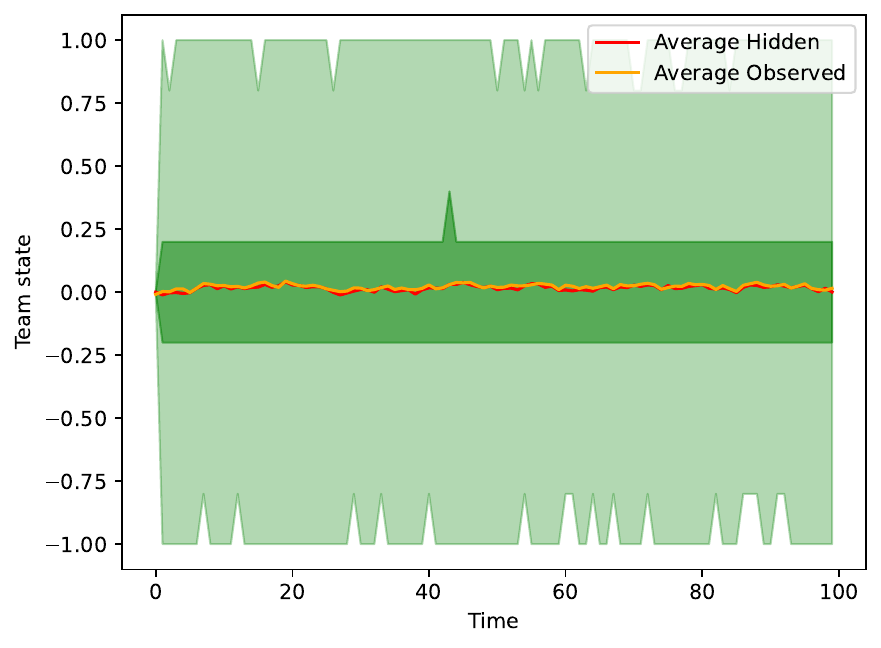}
        \caption{1Pillar + H + antiD}
    \end{subfigure}
    \hfill
    \begin{subfigure}[b]{\figwidth}
        \centering
    \includegraphics[width=\textwidth]{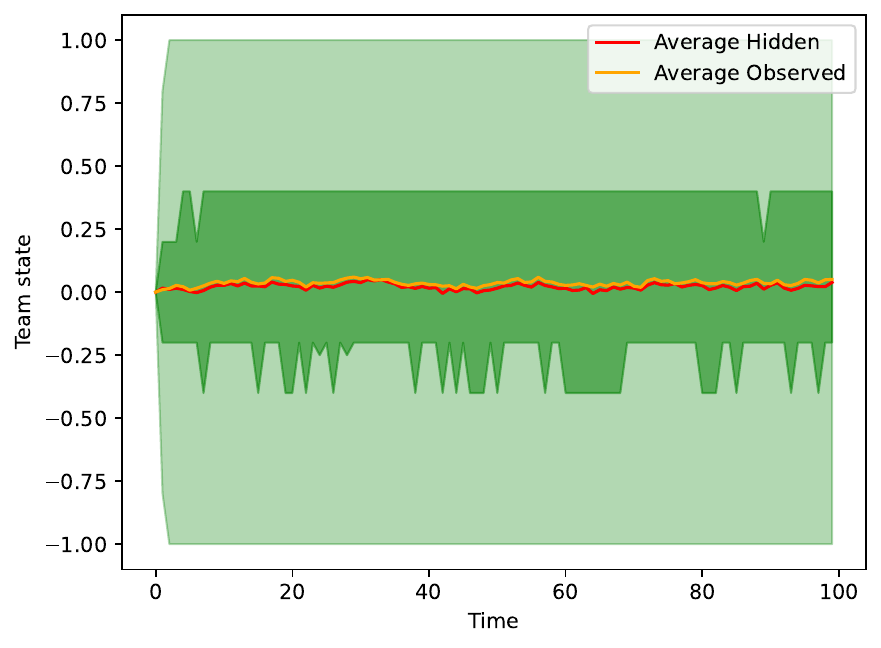}
        \caption{2Pillars}
    \end{subfigure}
    \begin{subfigure}[b]{\figwidth}
        \centering
     \includegraphics[width=\textwidth]{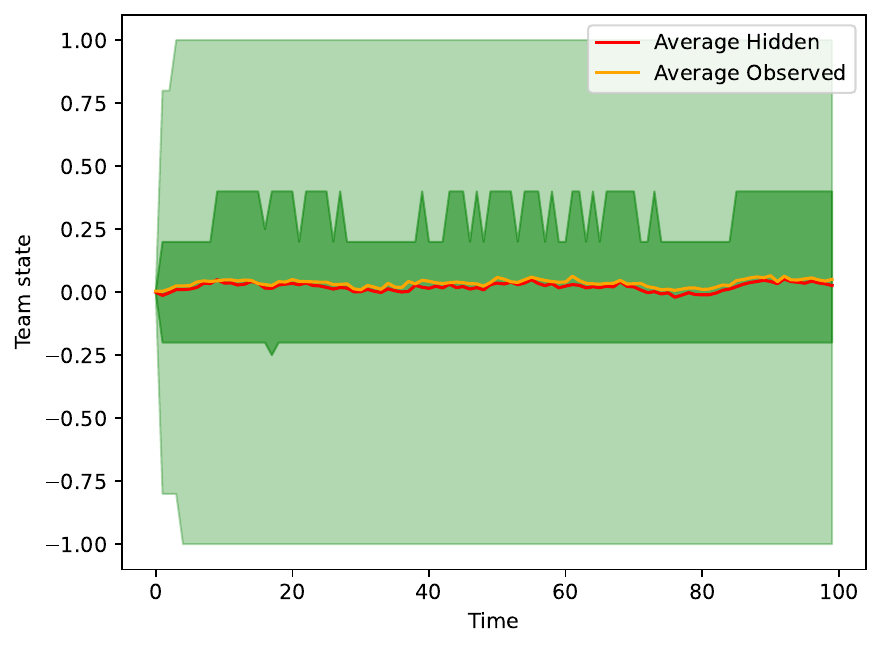}
        \caption{2Pillars + H}
    \end{subfigure}
    \hfill
    \begin{subfigure}[b]{\figwidth}
        \centering
    \includegraphics[width=\textwidth]{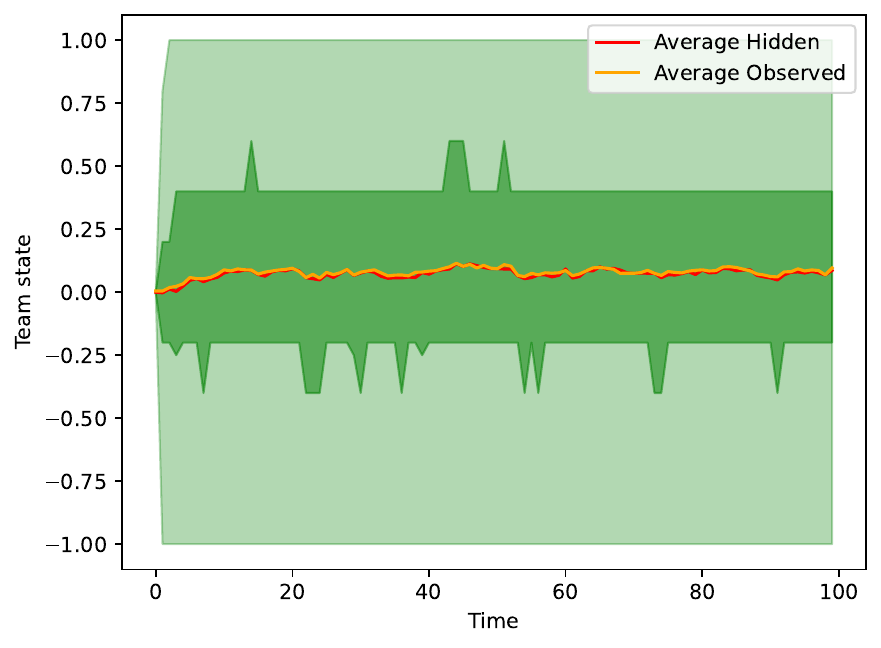}
        \caption{2Pillars + D}
    \end{subfigure}
    
    \caption{Average hidden and observed states for each team profile}
    \label{fig: team states (all teams) A}
\end{figure*}

\begin{figure*}[htbp]
    \centering
    \begin{subfigure}[b]{\figwidth}
        \centering
     \includegraphics[width=\textwidth]{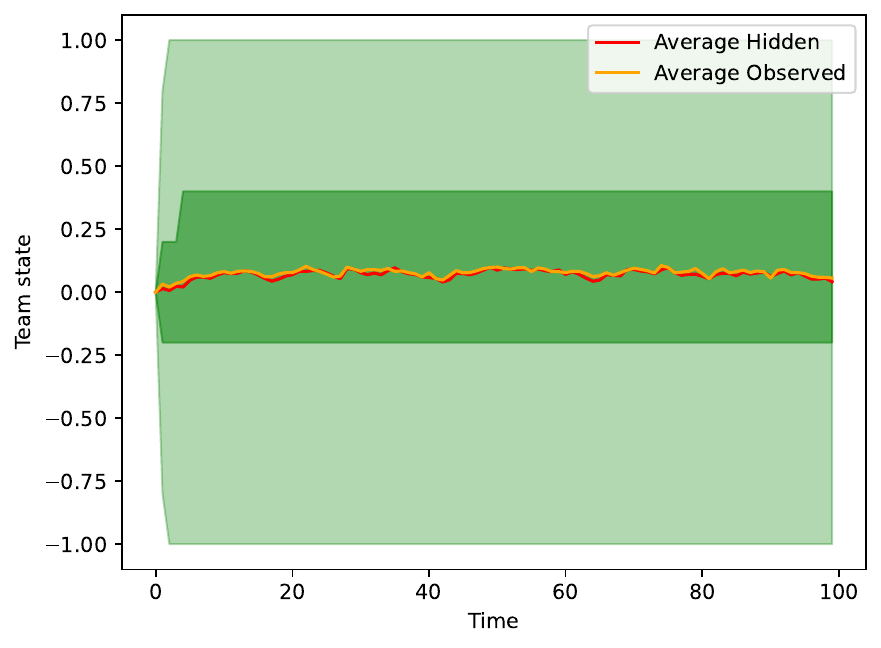}
        \caption{2Pillars + H + D}
    \end{subfigure}
    \hfill
    \begin{subfigure}[b]{\figwidth}
        \centering
    \includegraphics[width=\textwidth]{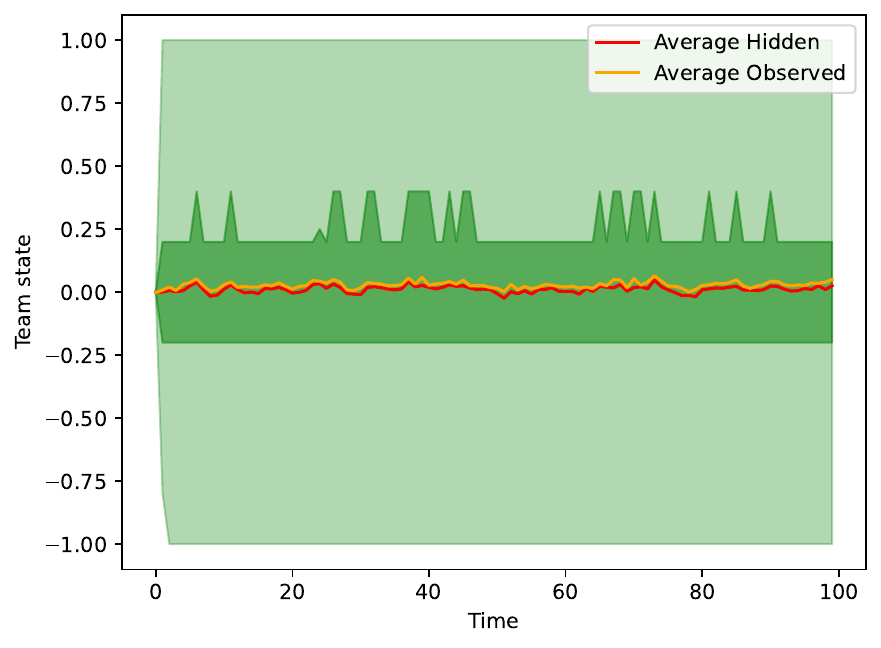}
        \caption{2Pillars + antiD}
    \end{subfigure}
    \begin{subfigure}[b]{\figwidth}
        \centering
    \includegraphics[width=\textwidth]{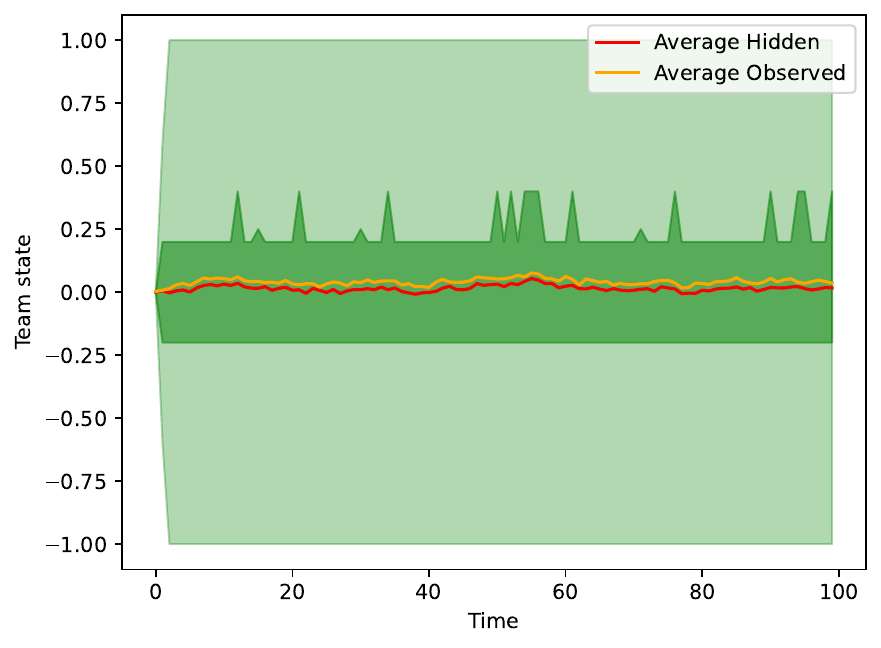}
        \caption{3Pillars}
    \end{subfigure}
    \hfill
    \begin{subfigure}[b]{\figwidth}
        \centering
    \includegraphics[width=\textwidth]{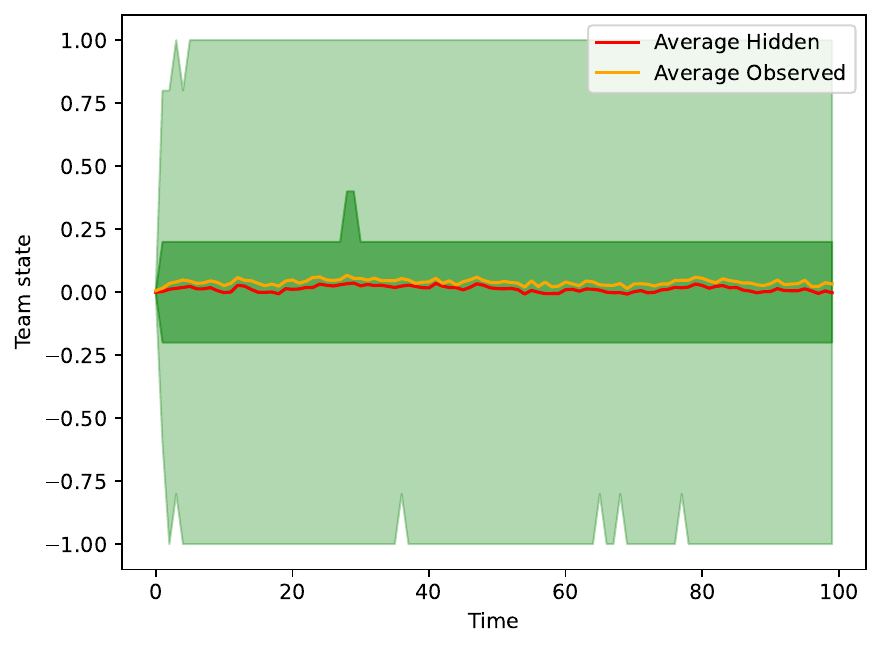}
        \caption{3Pillars + H}
    \end{subfigure}

    \begin{subfigure}[b]{\figwidth}
        \centering
     \includegraphics[width=\textwidth]{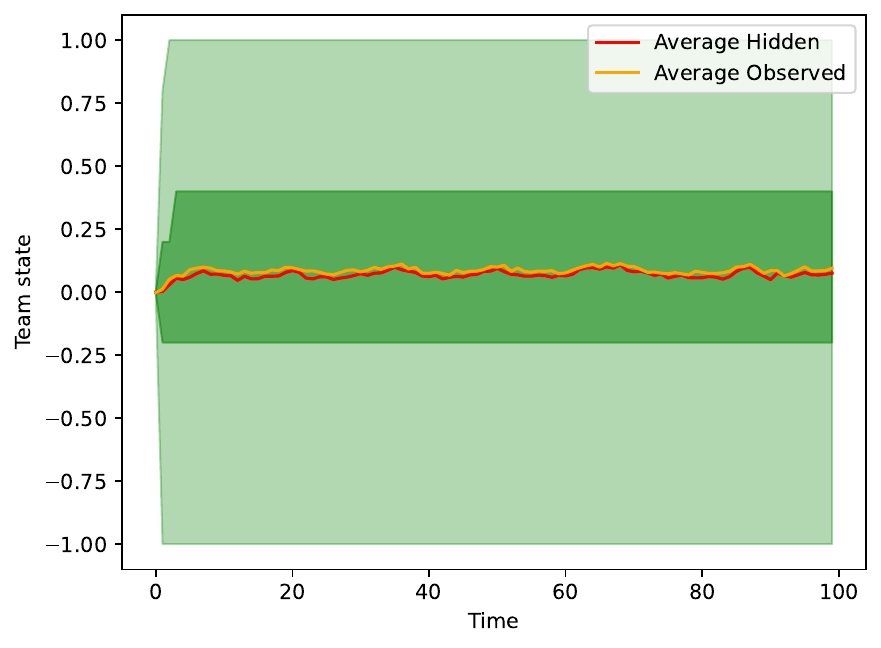}
        \caption{3Pillars + D}
    \end{subfigure}
    \hfill
    \begin{subfigure}[b]{\figwidth}
        \centering
    \includegraphics[width=\textwidth]{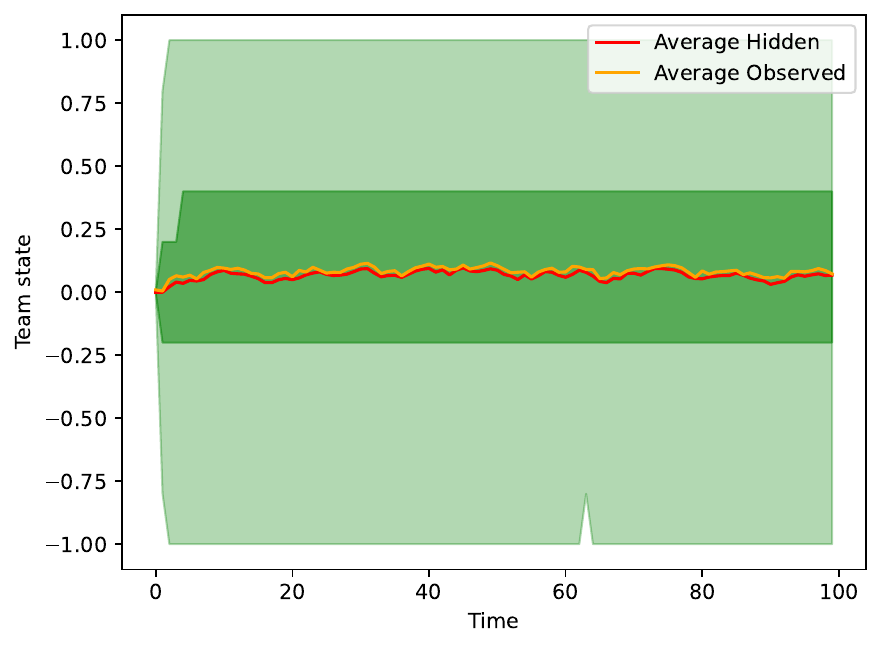}
        \caption{3Pillars + H + D}
    \end{subfigure}
    
    \caption{Average hidden and observed states for each team profile}
    \label{fig: team states (all teams) B}
\end{figure*}

\begin{figure*}[htbp]
    \centering
    \begin{subfigure}[b]{\figwidth}
        \centering
    \includegraphics[width=\textwidth]{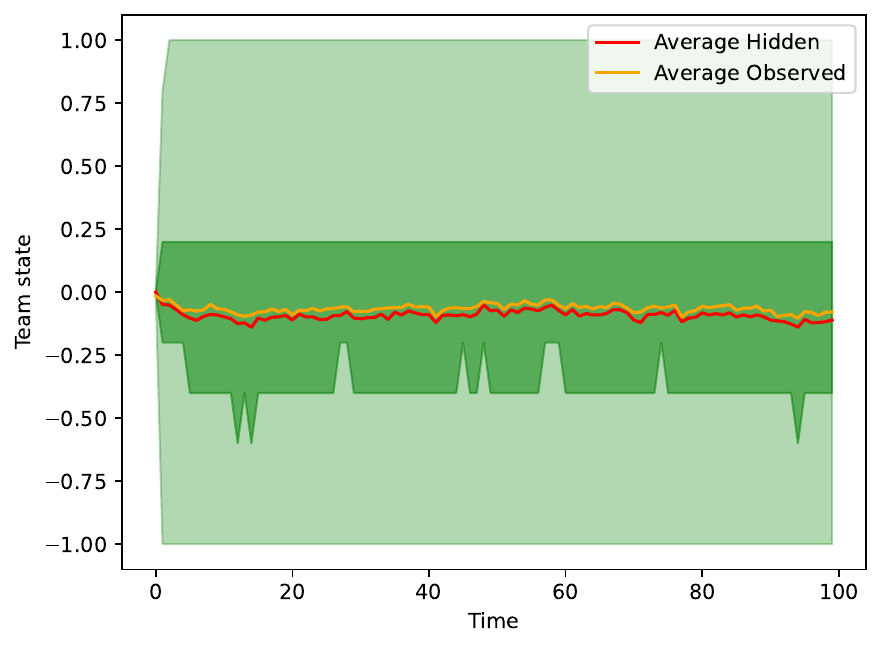}
        \caption{1Pillar Bad Teammate (BT1P)}
    \end{subfigure}
    \hfill
    \begin{subfigure}[b]{\figwidth}
        \centering
    \includegraphics[width=\textwidth]{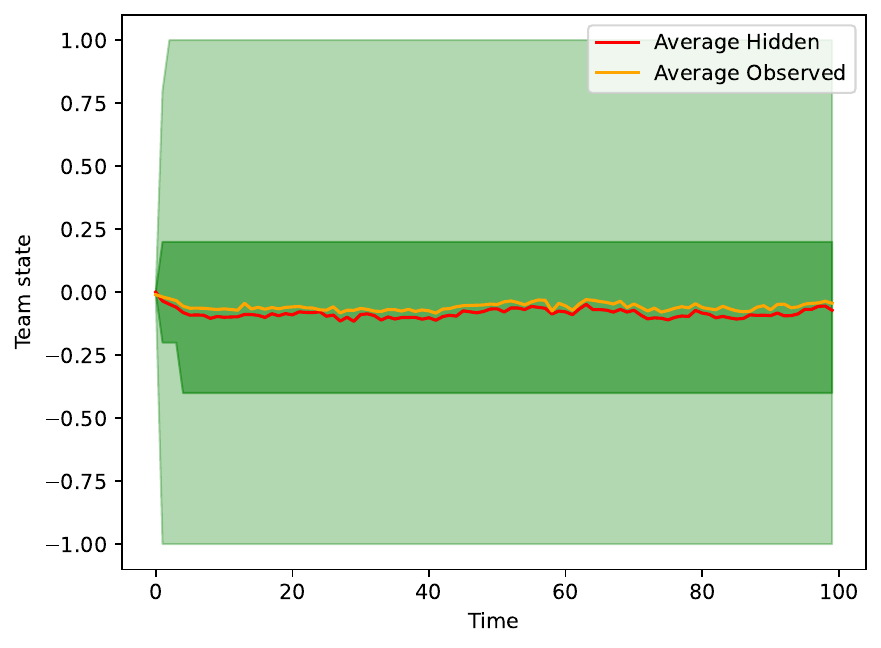}
        \caption{1Pillar Bad Teammate (BT1P) + H}
    \end{subfigure}

    \begin{subfigure}[b]{\figwidth}
        \centering
     \includegraphics[width=\textwidth]{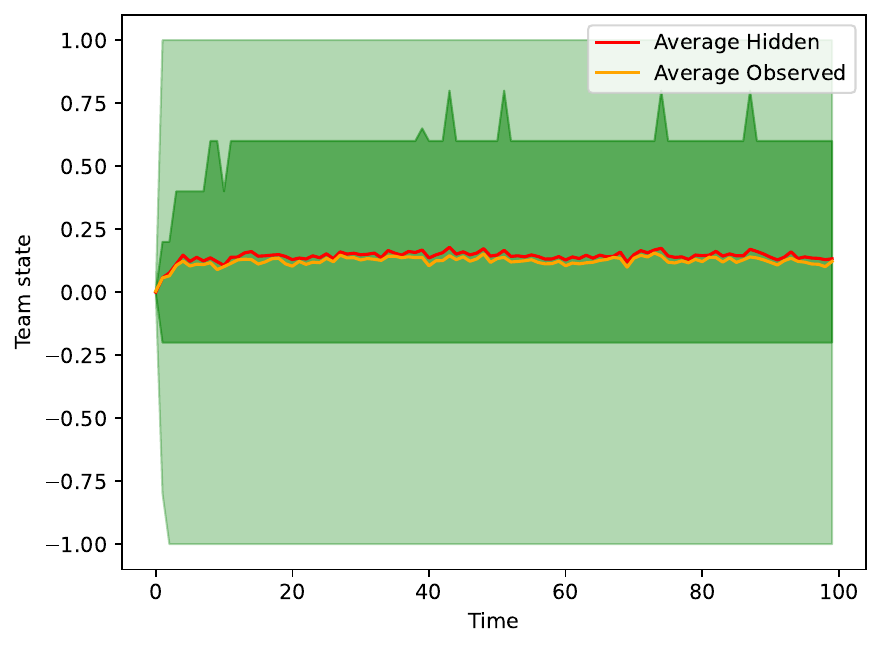}
        \caption{1Pillar Great Teammate (GT1T)}
    \end{subfigure}
    \hfill
    \begin{subfigure}[b]{\figwidth}
        \centering
    \includegraphics[width=\textwidth]{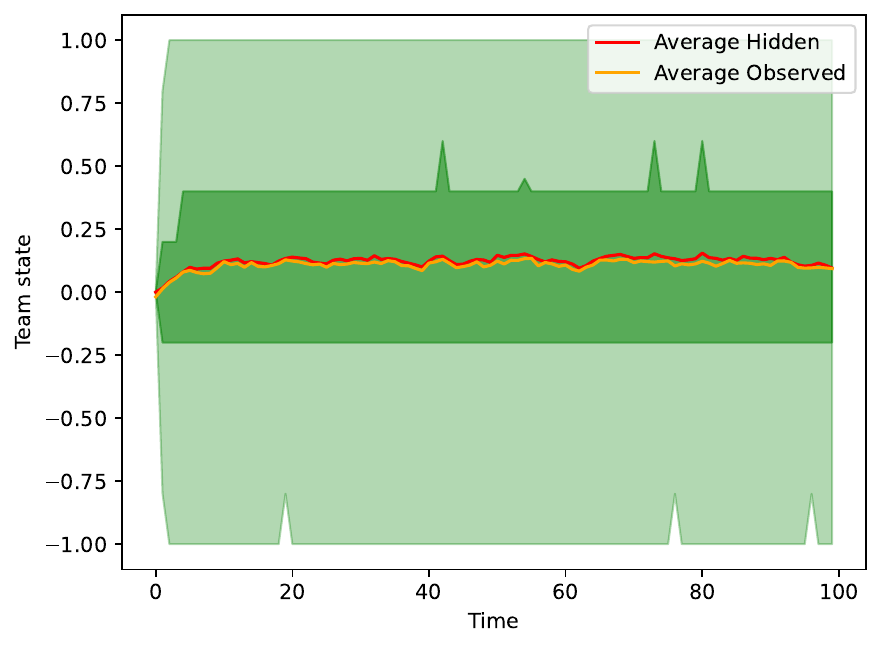}
        \caption{1Pillar Great Teammate (GT1T) + H}
    \end{subfigure}

    \begin{subfigure}[b]{\figwidth}
        \centering
     \includegraphics[width=\textwidth]{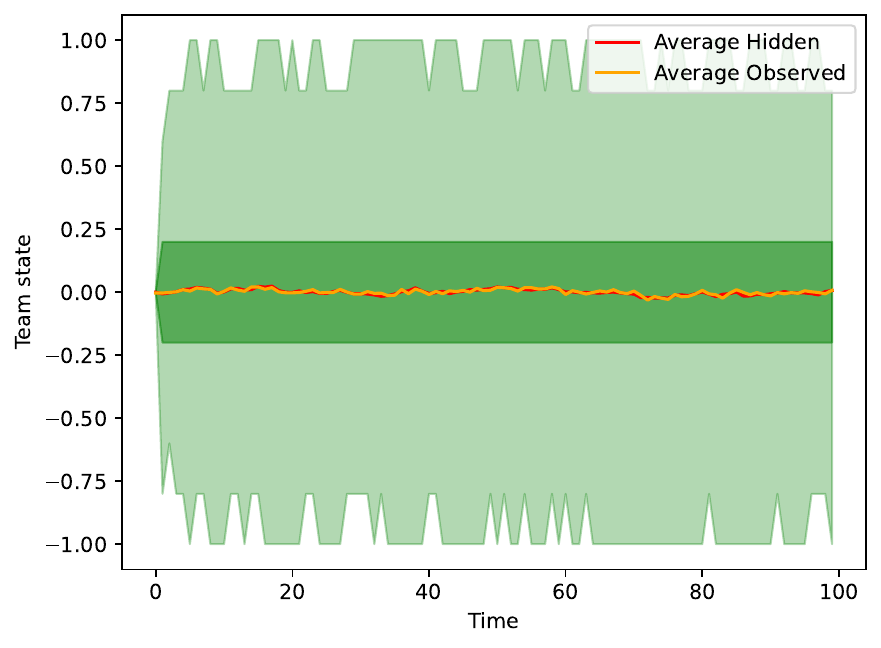}
        \caption{H}
    \end{subfigure}
    \hfill
    \begin{subfigure}[b]{\figwidth}
        \centering
    \includegraphics[width=\textwidth]{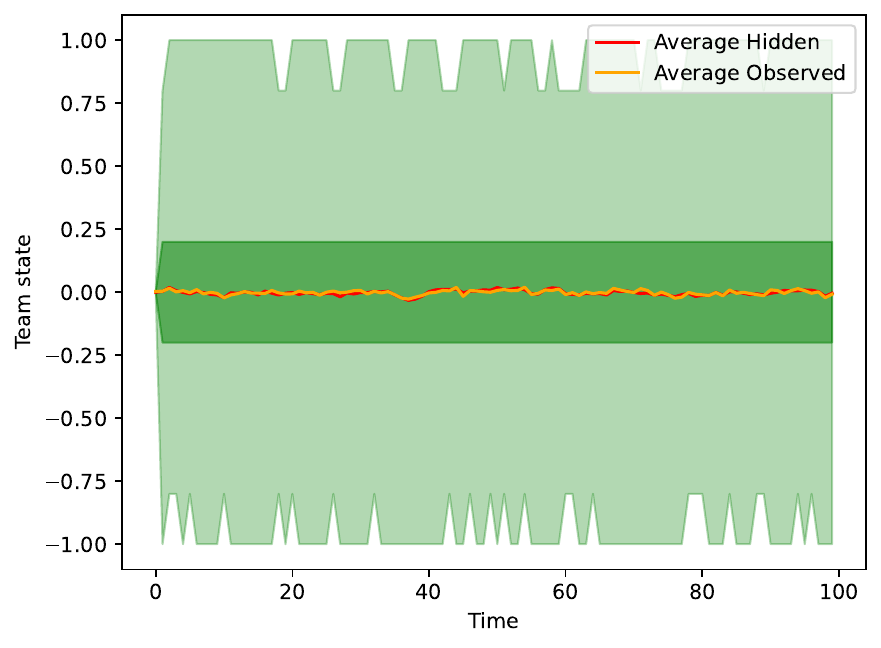}
        \caption{HO}
    \end{subfigure}
    
    \caption{Average hidden and observed states for each team profile}
    \label{fig: team states (all teams) C}
\end{figure*}

\begin{figure*}[htbp]
    \centering
    \begin{subfigure}[b]{\figwidth}
        \centering
     \includegraphics[width=\textwidth]{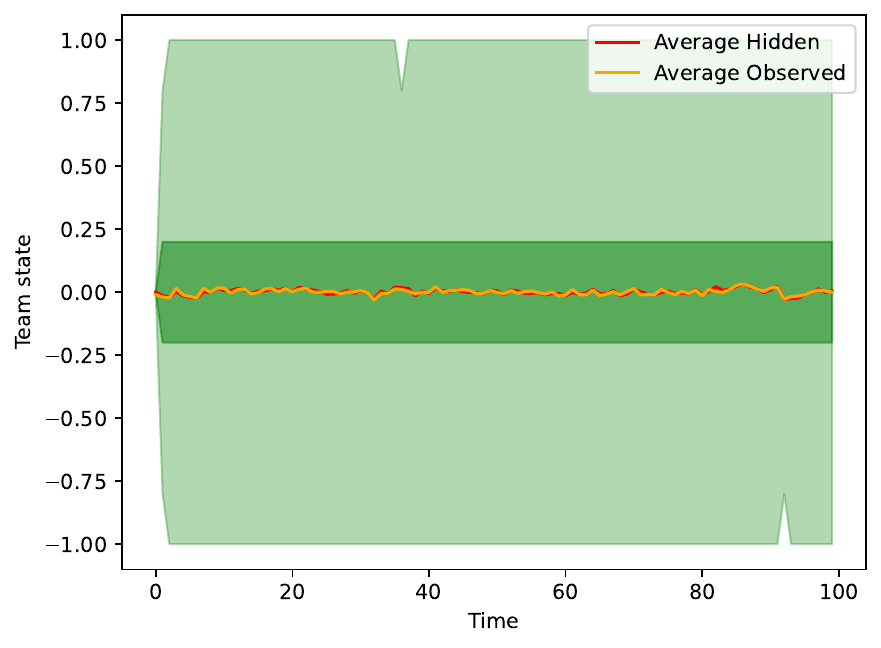}
        \caption{Balanced}
    \end{subfigure}
    
    \caption{Average hidden and observed team states for each team profile; average is taken over different simulations}
    \label{fig: team states (all teams) D}
\end{figure*}

We define the \emph{hidden team state} (resp.\ \emph{observed team state}) at each timestep as the average of the hidden (resp.\ observed) states of the individuals in the team at this timestep; values close to $0$ correspond to average team performance, values close to $-1$ (resp.\ $1$) correspond to an under-performing (resp.\ over-performing) team.
The team (hidden) state of the different teams over time is shown in Figures~\ref{fig: team states (all teams) A} - ~\ref{fig: team states (all teams) D}.
In these plots the thick green region corresponds to the 25-th and 75-th percentiles of 
the  team state across simulations.
The light green region corresponds to the min and max values of the team state across simulations.

Depending on the team profile, the thick green band might vary its width. Teams with a narrow green band, are more stable, since in $50\%$ of the time they are close to their average mental state. Furthermore, its worth noticing that for some profiles the green band is shifted either upwards (e.g. \texttt{GT1P}) or downwards (e.g \texttt{BT1P}). This shift reveals an inclination of a team towards higher (or lower) mental states.

\subsection{Team collapse probability vs Team profile} \label{app: team collapse probability vs profile}

\begin{figure}[H]
  \centering
\includegraphics[width=0.5\linewidth]{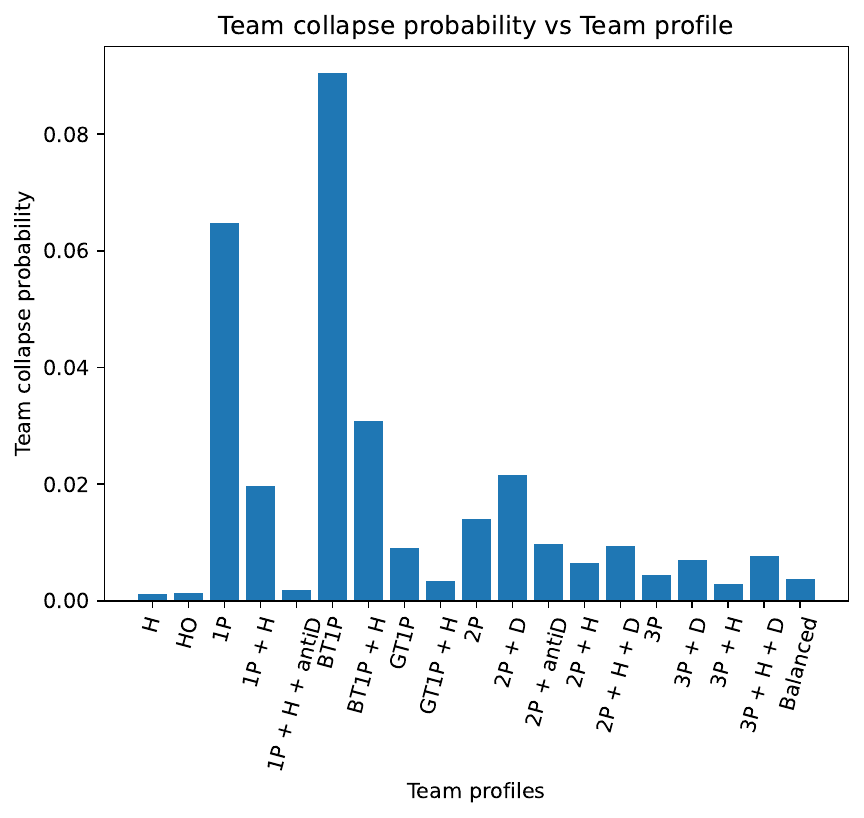}
  \caption{Team collapse probability vs Team profile}
  \Description{team collapse probability vs team profile}
  \label{fig: team_collapse_prob}
\end{figure}

\subsection{Maximum and average time in collapse state}

\begin{figure*}[htbp]
    \centering
    \begin{subfigure}[b]{\figwidth}
        \centering
     \includegraphics[width=\textwidth]{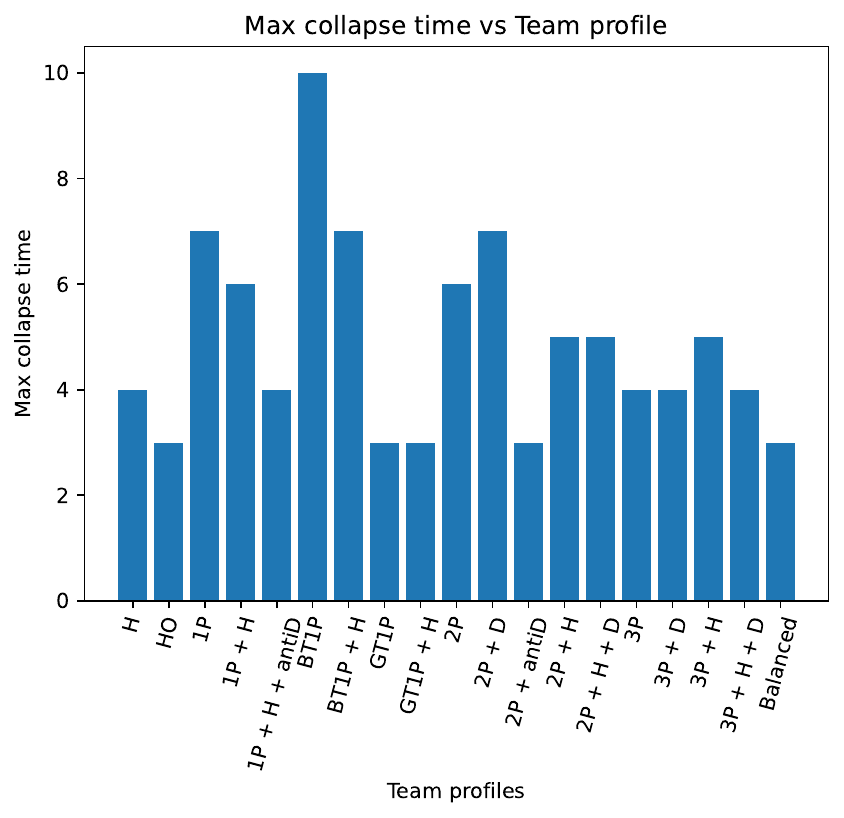}
        \caption{Maximum time in team collapse}
    \end{subfigure}
    \hfill
    \begin{subfigure}[b]{\figwidth}
        \centering
    \includegraphics[width=\textwidth]{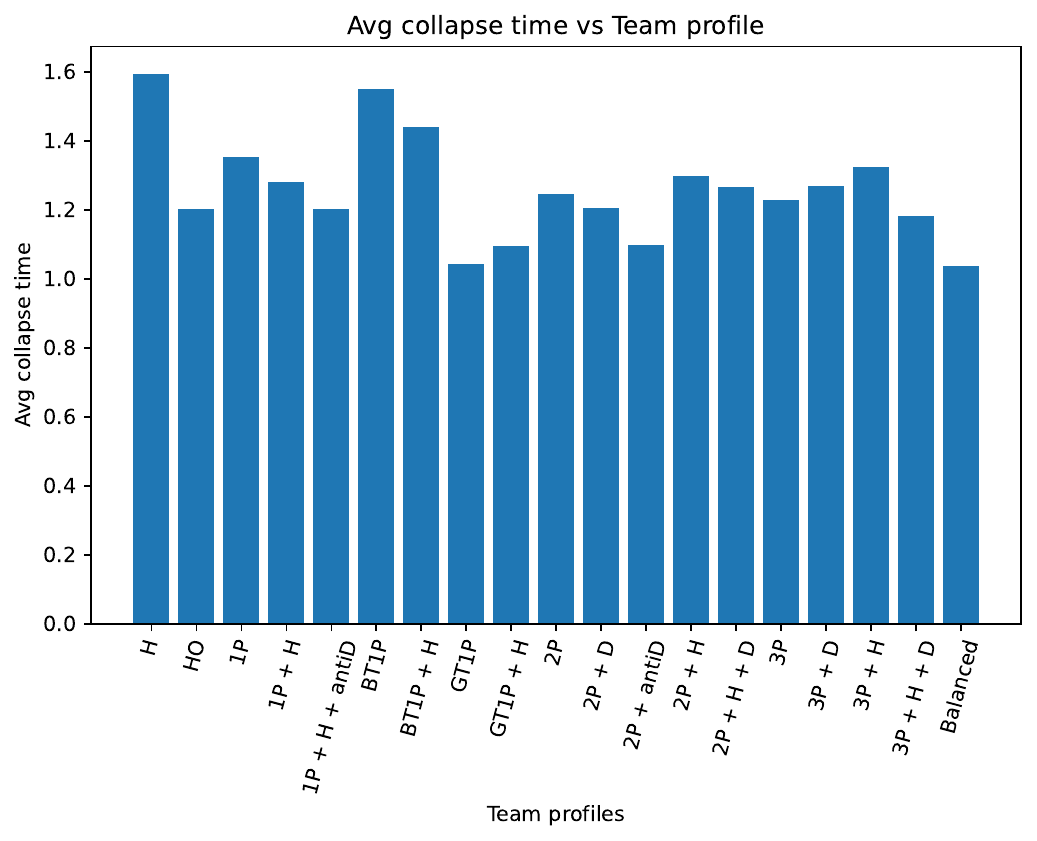}
        \caption{Average time in collapse}
    \end{subfigure}
    \caption{Maximum and average time team team collapse}
    \label{fig: max and avg time in collapse}
\end{figure*}

In figure \ref{fig: max and avg time in collapse} (a), we observe that the more pillars a team has, the longer the maximum window of team collapse is.
In general adding self-dependence, i.e. dependence on the hidden states, doesn't always make the team more resilient. However, adding self-dependence in the \texttt{3Pillars} profile resulted in higher maximum collapse time.
Moreover, we observe that having a Great Teammate player on whom all other players depend (\texttt{1Pillar Great Teammate (GT1T)} profile), significantly reduces the length of the maximum time collapse window. This is because when the pillar player under-performs, he/she only mildly affects non-pillar players. On the contrary, if the pillar player over-performs, non-pillar players are greatly affected in a positive way.

In Figure \ref{fig: max and avg time in collapse} (b), we observe that the average collapse time doesn't significantly vary among team profiles. 
That said, it is clear that when the players only depend on their hidden state, \ie \texttt{H} profile, the average duration of the collapse is longer. 
This is because, in the rare case that all players choke, it is very difficult to get out of the choke.
For example, this is not the case for the \texttt{1Pillar} profile.
When the pillar of the team chokes, non-pillar players are likely to choke. But, once the pillar player gets out of the choke, non-pillar players are likely to do the same.






\section{Independence}
In graphical models, independence between random variables is defined using the concept of $d$-separation. For a definition and analysis of $d$-separation see \cite{koller2009probabilistic}. We use $X \perp Y \mid Z$ to denote that $X$ is independent of $Y$ given $Z$.
In our proofs we used the following independence relationships:
\begin{itemize}
    \item $H_t^i \perp H_t^j \mid \Ob_{1:t}$
    \item $O_t^i \perp \Ob_t^{-i} \mid H_t^i$
    \item $H_t^i \perp \Ob_t^{-i} \mid \Ob_{1:t-1}$
    \item $O_t^i \perp \Ob_t^{1:t-1} \mid H_t^i$
    \item $(H_t^i, O_t^i) \perp (H_t^j, O_t^j)  \mid \Ob_{1:t-1}$
    \item $\Hb^i \perp \Hb^j \mid \Ob$
\end{itemize}

Apart from checking the above relationships using the rules of $d$-separation, we also checked them using independence checker programs that are freely available online.

\section{Appendix for Section~\ref{sec: learning}}\label{app:em}
\subsection{Expectation step}~\label{app:expectation}
\spara{Calculating the posterior.} Using the definition of conditional probability we have that
the posterior can be written as
\begin{align}\label{eq:posterior}
    \Prob(\Hb \mid \Ob) =
    \frac{\Prob(\Hb, \Ob)}{\Prob(\Ob)}.
\end{align}
Now, using the structure of the graphical model (see Fig.~\ref{fig: graphical_model}), we can decompose the joint probability of the variables  as a product of probabilities where each variable is conditioned on its parent nodes~\cite{koller2009probabilistic}. Thus, we have
\begin{align} \label{eq:numerator}
    \Prob(\Hb, \Ob) =
    \Prob(\Hb_1)
    \prod_{t=1}^T \prod_{i=1}^n \Prob(O_t^i \mid H_t^i)
    \prod_{t=2}^T \prod_{i=2}^n \Prob(H_t^i \mid H_{t-1}^i, \Ob_{t-1})
\end{align}

Thus, the difficulty of sampling from the posterior stems from the denominator $\Prob(\Ob)$ which can be written as
\begin{align}
    \Prob(\Ob) = \Prob(\Ob_1) \prod \Prob(\Ob_t \mid \Ob_{1:t-1}).
\end{align}
In general, calculating $\Prob(\Ob)$ is difficult, however in our model it can be done efficiently
by calculating $\Prob(\Ob_t \mid \Ob_{1:t-1})$. We do this as follows.
First, we define the forward parameters
\begin{align}
\label{eq: forward-def}
    \Prob({\Hb_t \mid \Ob_{1:t}}) = 
    \prod_{i=1}^n \Prob(H_t^i \mid \Ob_{1:t}),
\end{align}
where we used the independence of $H_t^i$ and $H_t^j$ given $\Ob_{1:t}$. We now calculate $\Prob(H_t^i | \Ob_{1:t-1})$.
\begin{align}
\Prob(H_t^i \mid \Ob_{1:t}) 
    &= \Prob(H_t^i \mid \Ob_{1:t-1}, \Ob_t)\\
    &= \frac{\Prob(\Ob_t \mid H_t^i, \Ob_{1:t-1}) \Prob(H_t^i \mid \Ob_{1:t-1})}{\Prob(\Ob_t \mid \Ob_{1:t-1})}\\
    &= \frac{\Prob(\Ob_t^{-i} \mid \Ob_{1:t-1}) \Prob(O_t^i \mid H_t^i, \Ob_{1:t-1}) \Prob(H_t^i \mid \Ob_{1:t-1})}
    {\Prob(O_t^i \mid \Ob_{1:t-1}) \Prob(\Ob_t^{-i} \mid \Ob_{1:t-1})}\\
    &= \frac{\Prob(O_t^i \mid H_t^i) \Prob(H_t^i \mid \Ob_{1:t-1})}
    {\Prob(O_t^{i} \mid \Ob_{1:t-1})},
\end{align}
where in the second equation we used Bayes' rule, in the third equation we used the independence between $O_t^i$ and $\Ob_t^{-i}$ given $\Ob_{1:t-1}$ and the independence of $\Ob_t^{-i}$ and $H_t^i$ given $\Ob_{1:t-1}$, and in the last equation we used the independence of $O_t^i$ and $\Ob_{1:t-1}$ given $H_t^i$.
Furthermore, we have
\begin{align}
    \Prob(H_t^i \mid \Ob_{1:t-1}) = 
    \sum_{h \in \mathcal{H}} \Prob(H_t^i \mid H_{t-1}^i = h, \Ob_{1:t-1}) \Prob(H_{t-1}^i \mid \Ob_{1:t-1}),
\end{align}
where $\Prob(H_{t-1}^i \mid \Ob_{1:t-1})$ can be calculated recursively. 

We also have
\begin{align}
\Prob(\Hb_t, \Ob_t \mid \Ob_{1:t-1}) 
&=
\prod_{i=1}^n \Prob(H_t^i, O_t^i \mid \Ob_{1:t-1}) \\
&=
\prod_{i=1}^n \Prob(O_t^i \mid H_t^i, \Ob_{1:t-1}) \Prob(H_t^i \mid \Ob_{1:t-1})\\
&=
\prod_{i=1}^n \Prob(O_t^i \mid H_t^i) \Prob(H_t^i \mid \Ob_{1:t-1})
,
\end{align}
where in the first equation we used the independence of $(H_t^i, O_t^i)$ and $(H_t^j, O_t^j)$ given $\Ob_{1:t-1}$, and in the third equation we used the independence of $O_t^i$ and $\Ob_{1:t-1}$ given $H_t^i$.

Moreover, we have
\begin{align}
\label{eq: norm-const}
    \Prob(\Hb_t \mid \Ob_{1:t}) &=
    \frac{\Prob(\Hb_t, \Ob_t \mid \Ob_{1:t-1})}{\Prob(\Ob_t \mid \Ob_{1:t-1})}
\end{align}

Having computed $\Prob(\Hb_t \mid \Ob_{1:t})$ and $\Prob(\Hb_t, \Ob_t \mid \Ob_{1:t-1})$ we can calculate $\Prob(\Ob_t \mid \Ob_{1:t-1})$ using the above equation.
Thus, we can compute $\Prob(\Ob)$ and the posterior distribution.

The E-step concludes by defining 
$Q(\params \mid \params^\text{old})$ as the expected value of the log likelihood function of 
$\params$, with respect to the current conditional distribution of 
$\Hb$ given 
$\Ob$ and the current estimates of the parameters:
\begin{align}
    Q(\params \mid \params^\text{old}) = 
    \EX_{\Hb \sim \Prob(\Hb \mid \Ob, \params^\text{old})}
    \left[
    \log \Prob(\Hb, \Ob \mid \params)
    \right]
\end{align}

In practice, we estimate the expectation by drawing samples from the posterior distribution $\Prob(\Hb \mid \Ob, \params^\text{old})$, and using the empirical mean. To this end, we next show how to generate samples from the posterior distribution.

\spara{Sampling from the posterior.}
We are now ready to prove Lemma~\ref{lemma:sampling} that allows us to
 decompose the posterior distribution into products of simpler distributions. 
\begin{proof}
    First, based on the previous analysis, we have
    \begin{align}
        \Prob(\Ob_t \mid \Ob_{1:t-1}) &= \frac{\Prob(\Hb_t, \Ob_t \mid \Ob_{1:t-1})}{\Prob(\Hb_t \mid \Ob_{1:t})} \\
        &= \prod_{i=1}^n \frac{\Prob(H_t^i, O_t^i \mid \Ob_{1:t-1})}{\Prob(H_t^i \mid \Ob_{1:t})} \\
        &= \prod_{i=1}^n \frac{\Prob(O_t^i \mid H_t^i) \Prob(H_t^i \mid \Ob_{1:t-1})}{\Prob(H_t^i \mid \Ob_{1:t})}.
    \end{align}
    Hence, the evidence becomes
    \begin{align}
        \Prob(\Ob) &= \prod_{t=1}^T \Prob(\Ob_t \mid \Ob_{1:t-1}) \\
        &= \prod_{t=1}^T \prod_{i=1}^n \frac{\Prob(H_t^i, O_t^i \mid \Ob_{1:t-1})}{\Prob(H_t^i \mid \Ob_{1:t})} \\
        &= \prod_{t=1}^T \prod_{i=1}^n \frac{\Prob(O_t^i \mid H_t^i) \Prob(H_t^i \mid \Ob_{1:t-1})}{\Prob(H_t^i \mid \Ob_{1:t})}.
    \end{align}
    Finally, substituting the above into the definition of the posterior distribution and combining with the joint distribution, we have
    \begin{align}
        Pr(\Hb \mid \Ob) &= \frac{\Prob(\Hb, \Ob)}{\Prob(\Ob)} \\
        &= 
    \prod_{t=1}^T \prod_{i=1}^n 
    \frac{\Prob(O_t^i \mid H_t^i)
    \Prob(H_t^i \mid H_{t-1}^i, \Ob_{t-1})}{\Prob(O_t^i \mid H_t^i) \Prob(H_t^i \mid \Ob_{1:t-1}) / \Prob(H_t^i \mid \Ob_{1:t})
    } \\
    &= \prod_{t=1}^T \prod_{i=1}^n 
    \frac{\Prob(H_t^i \mid \Ob_{1:t})
    \Prob(H_t^i \mid H_{t-1}^i, \Ob_{t-1})}{ \Prob(H_t^i \mid \Ob_{1:t-1})
    }
    \end{align}.
\end{proof}

Thus, we sample $H_t^i$ from a propability distribution that is proportional to
\begin{align}
  \frac{\Prob(H_t^i = h \mid \Ob_{1:t})
    \Prob(H_t^i = h\mid H_{t-1}^i, \Ob_{t-1})}{ \Prob(H_t^i = h \mid \Ob_{1:t-1})}.
\end{align}

\section{Real-world experiments}

\subsection{Team structures}

\begin{figure*}[htbp]
    \centering
    \begin{subfigure}[b]{\figwidth}
        \centering
     \includegraphics[width=\textwidth]{img/heatmap/BOS_heatmap.pdf}
        \caption{Boston Celtics}
    \end{subfigure}
    \hfill
    \begin{subfigure}[b]{\figwidth}
        \centering
    \includegraphics[width=\textwidth]{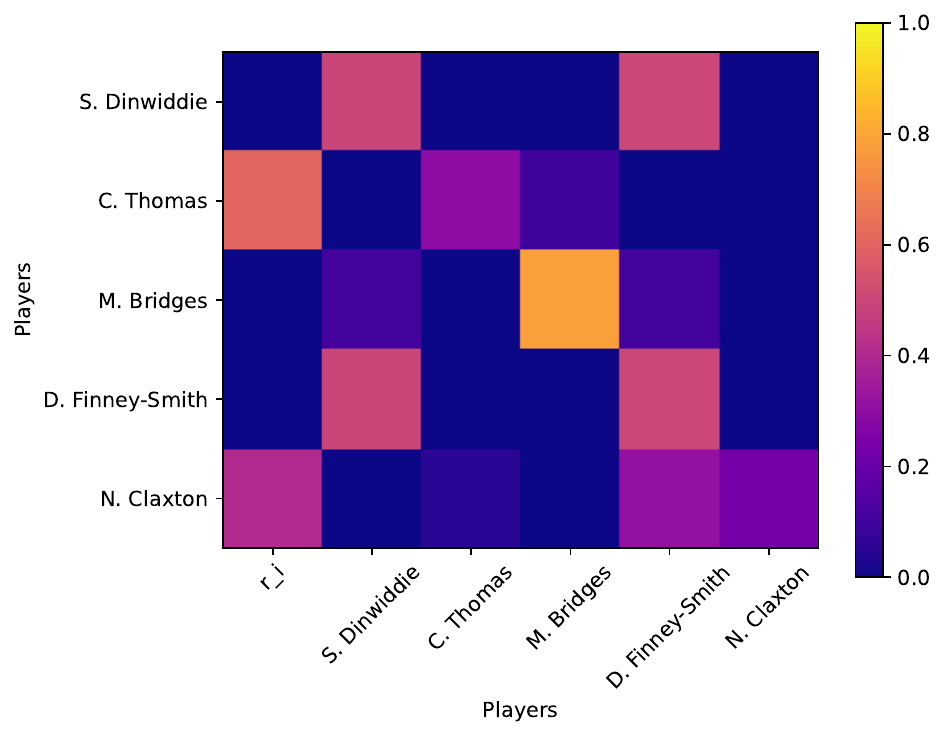}
        \caption{Brooklyn Nets}
    \end{subfigure}

    \begin{subfigure}[b]{\figwidth}
        \centering
    \includegraphics[width=\textwidth]{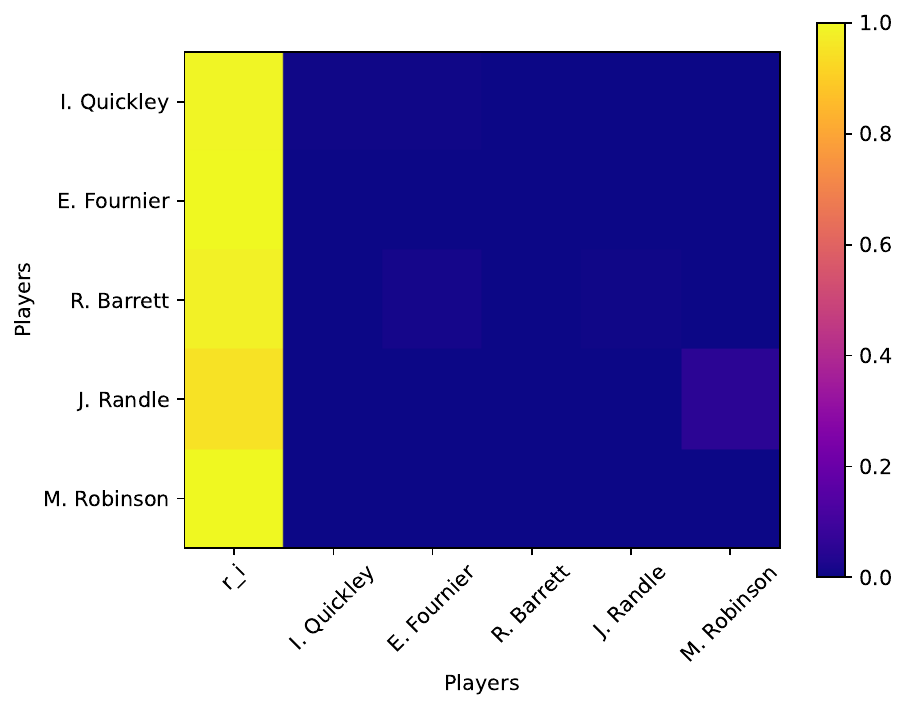}
        \caption{New York Knicks}
    \end{subfigure}
    \hfill
    \begin{subfigure}[b]{\figwidth}
        \centering
    \includegraphics[width=\textwidth]{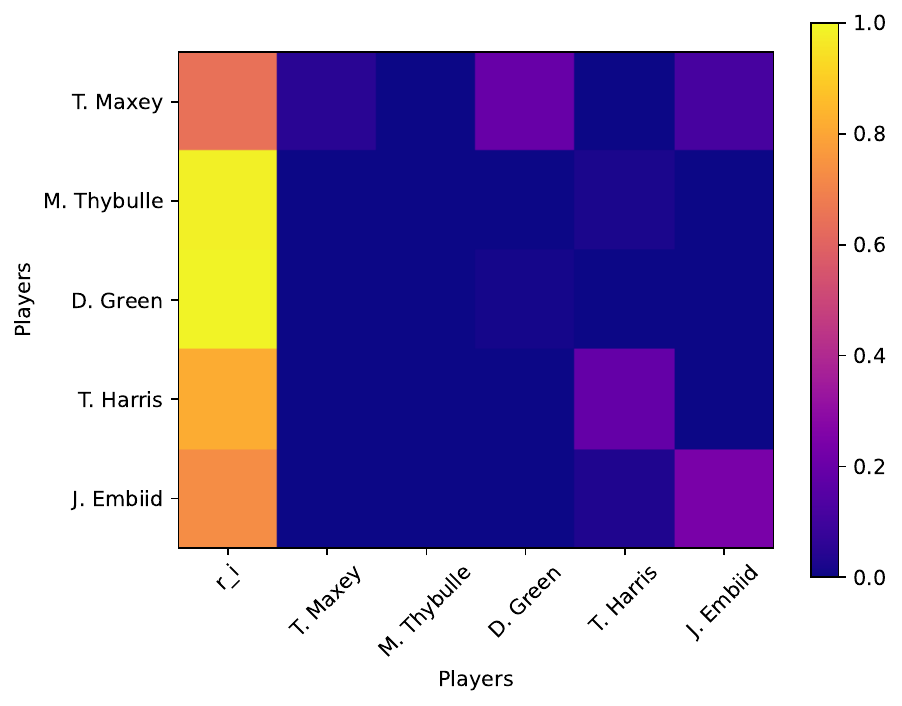}
        \caption{Philadelphia 76ers}
    \end{subfigure}
    \begin{subfigure}[b]{\figwidth}
        \centering
    \includegraphics[width=\textwidth]{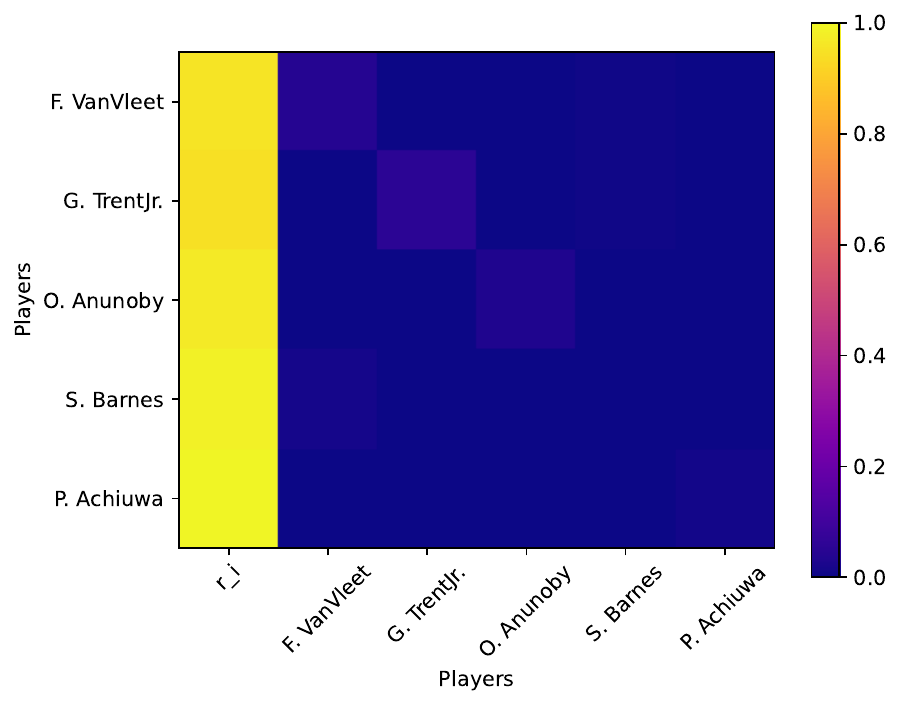}
        \caption{Toronto Raptors}
    \end{subfigure}
    \hfill
    \begin{subfigure}[b]{\figwidth}
        \centering
    \includegraphics[width=\textwidth]{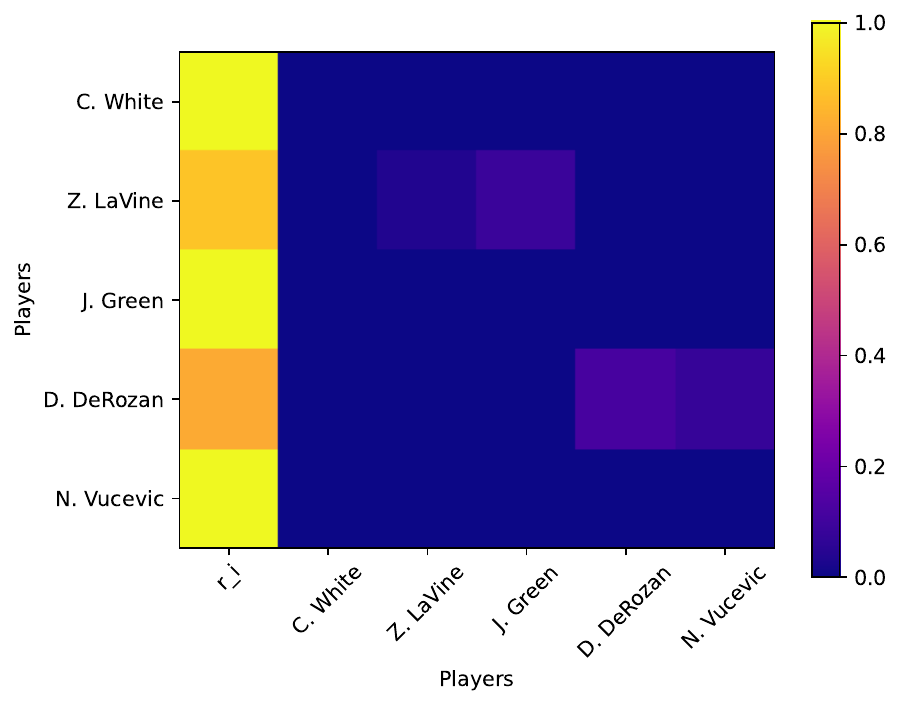}
        \caption{Chicago Bulls}
    \end{subfigure}
    \caption{Team structure $\Rmatrix$}
    \label{fig: team-structure A}
\end{figure*}

Figure \ref{fig: team-structure A} presents the structure matrix for all 30 NBA teams in our dataset. 
As discussed in Section \ref{sec: real-world experiments} there are two main profiles and each team can be thought of as falling closer to one of them. For example the Cleveland Cavaliers, fall closer to the profile we saw for the Dallas Mavericks, but with the added difference that two of their players also depend slightly to their own observed state (not only their hidden state). 
At the same time, Houston Rockets are closer to the profile we saw for the Boston Celtics, where the observed state of Martin seems to impact the state of his teammates. 

\label{app: team-profiles}

\begin{figure*}[htbp]
    \centering
    \begin{subfigure}[b]{\figwidth}
        \centering
     \includegraphics[width=\textwidth]{img/heatmap/CLE_heatmap.pdf}
        \caption{Cleveland Cavaliers}
    \end{subfigure}
    \hfill
    \begin{subfigure}[b]{\figwidth}
        \centering
    \includegraphics[width=\textwidth]{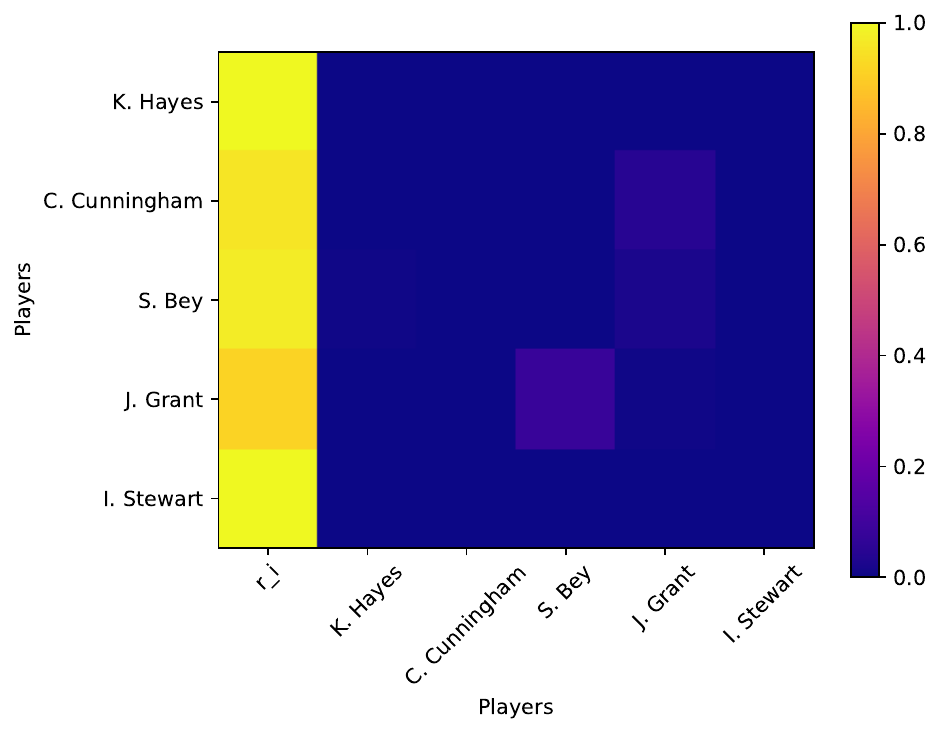}
        \caption{Detroit Pistons}
    \end{subfigure}

    \begin{subfigure}[b]{\figwidth}
        \centering
    \includegraphics[width=\textwidth]{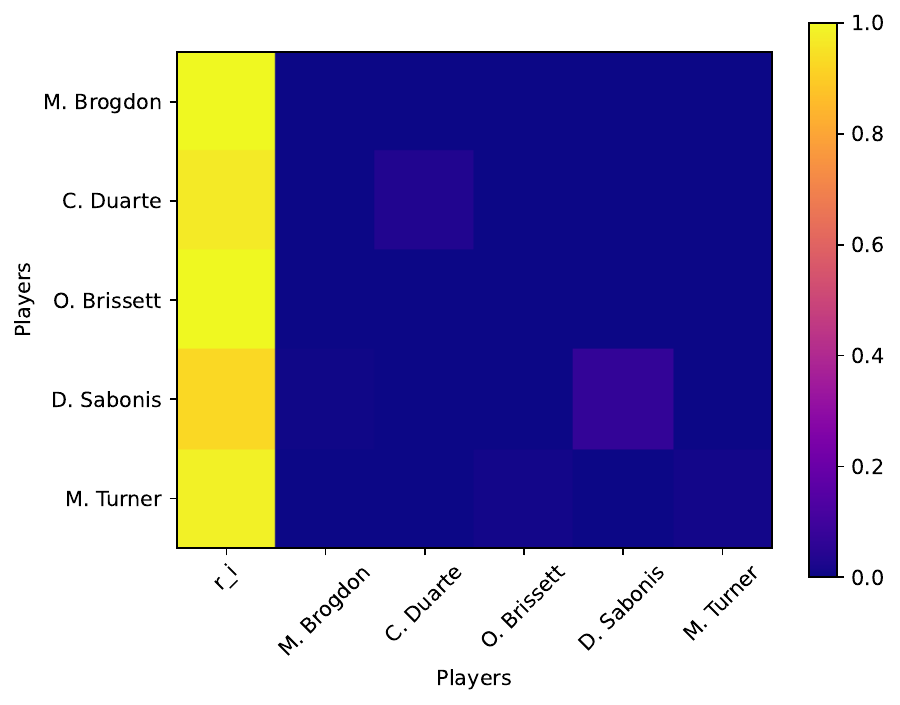}
        \caption{Indiana Pacers}
    \end{subfigure}
    \hfill
    \begin{subfigure}[b]{\figwidth}
        \centering
    \includegraphics[width=\textwidth]{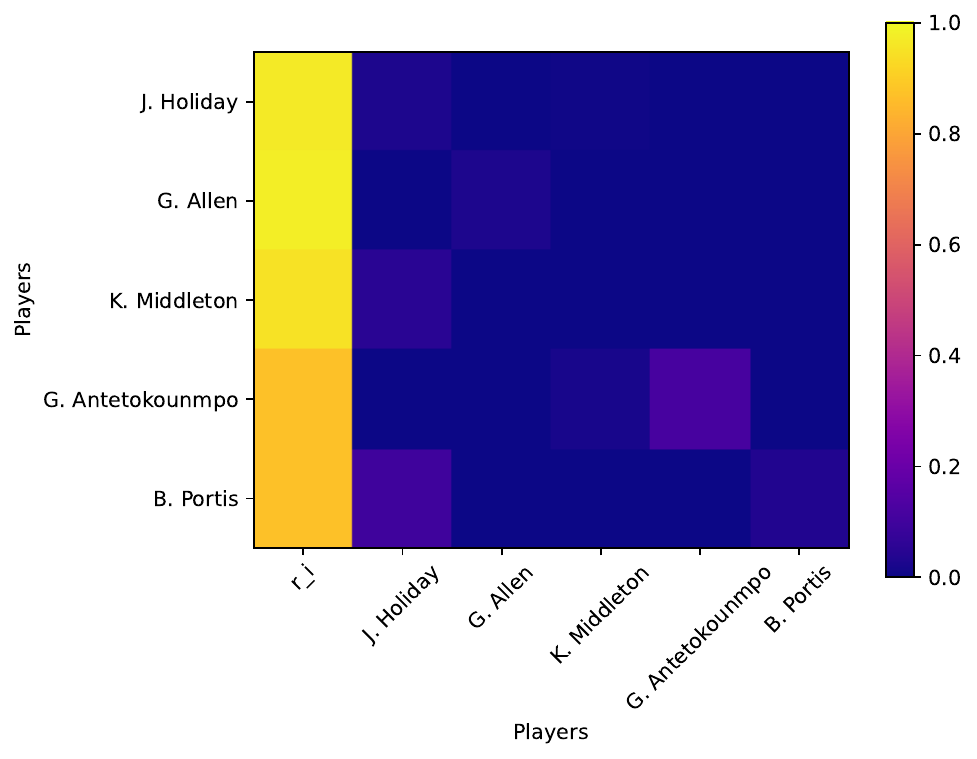}
        \caption{Milwaukee Bucks}
    \end{subfigure}
    \begin{subfigure}[b]{\figwidth}
        \centering
    \includegraphics[width=\textwidth]{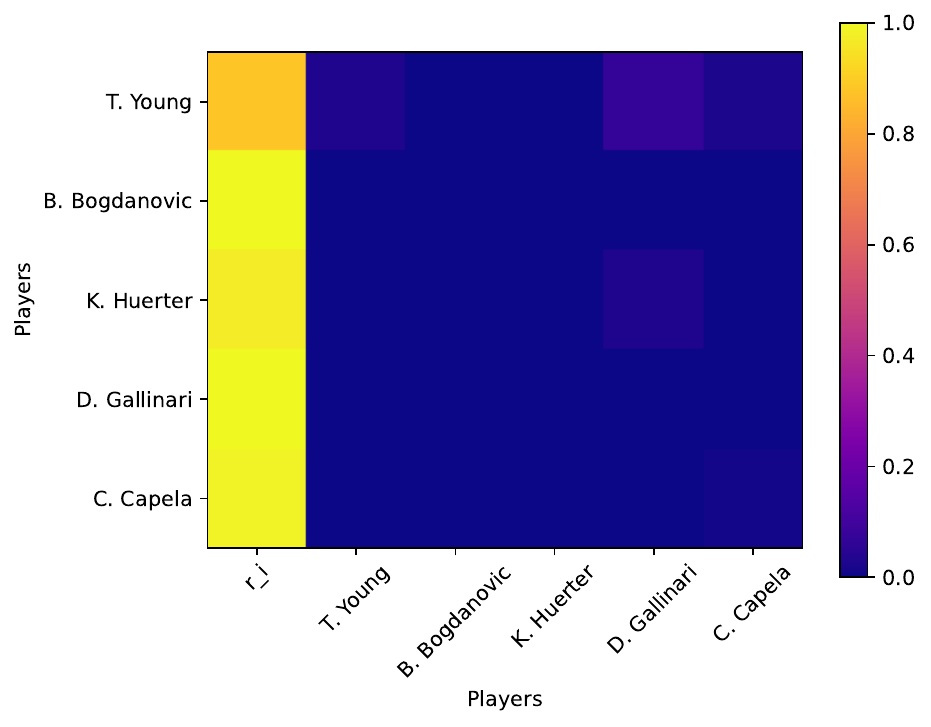}
        \caption{Atlanta Hawks}
    \end{subfigure}
    \hfill
    \begin{subfigure}[b]{\figwidth}
        \centering
    \includegraphics[width=\textwidth]{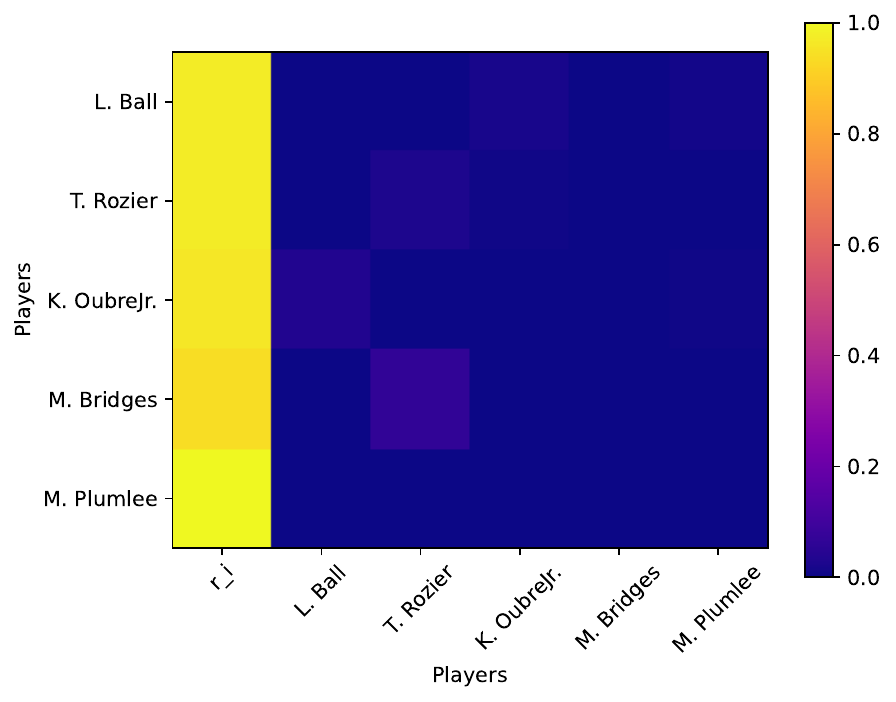}
        \caption{Charlotte Hornets}
    \end{subfigure}
    \caption{Team structure $\Rmatrix$}
    \label{fig: team-structure B}
\end{figure*}

\begin{figure*}[htbp]
    \centering
    \begin{subfigure}[b]{\figwidth}
        \centering
     \includegraphics[width=\textwidth]{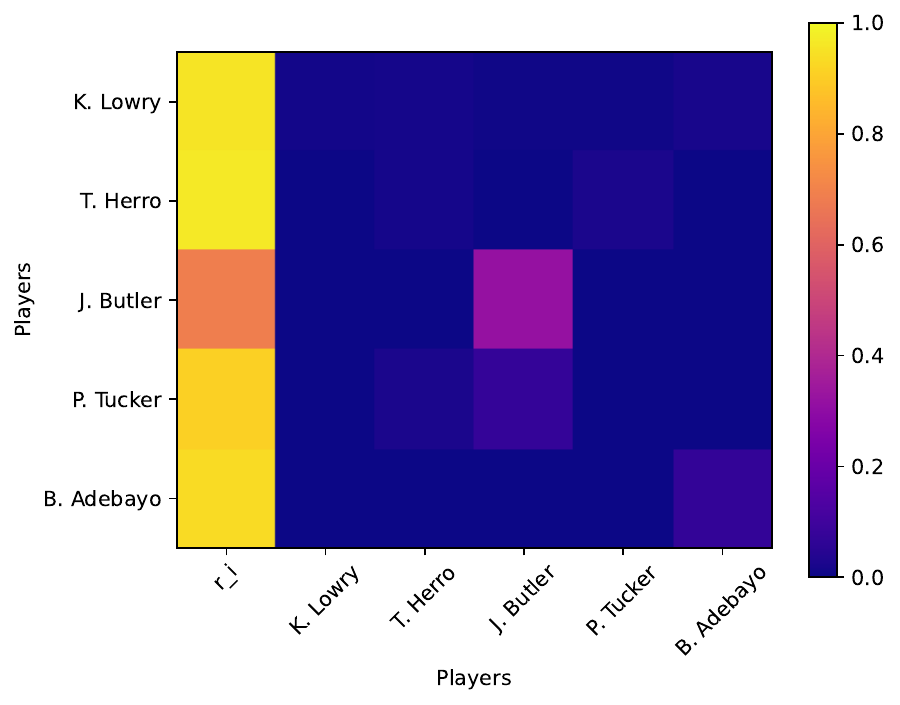}
        \caption{Miami Heat}
    \end{subfigure}
    \hfill
    \begin{subfigure}[b]{\figwidth}
        \centering
    \includegraphics[width=\textwidth]{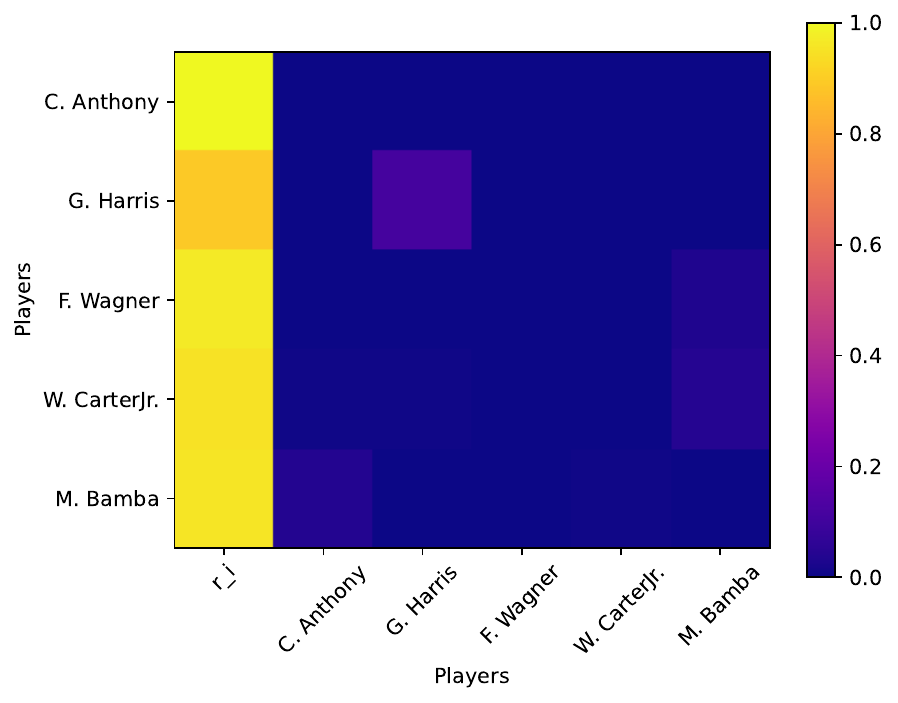}
        \caption{Orlando Magic}
    \end{subfigure}

    \begin{subfigure}[b]{\figwidth}
        \centering
    \includegraphics[width=\textwidth]{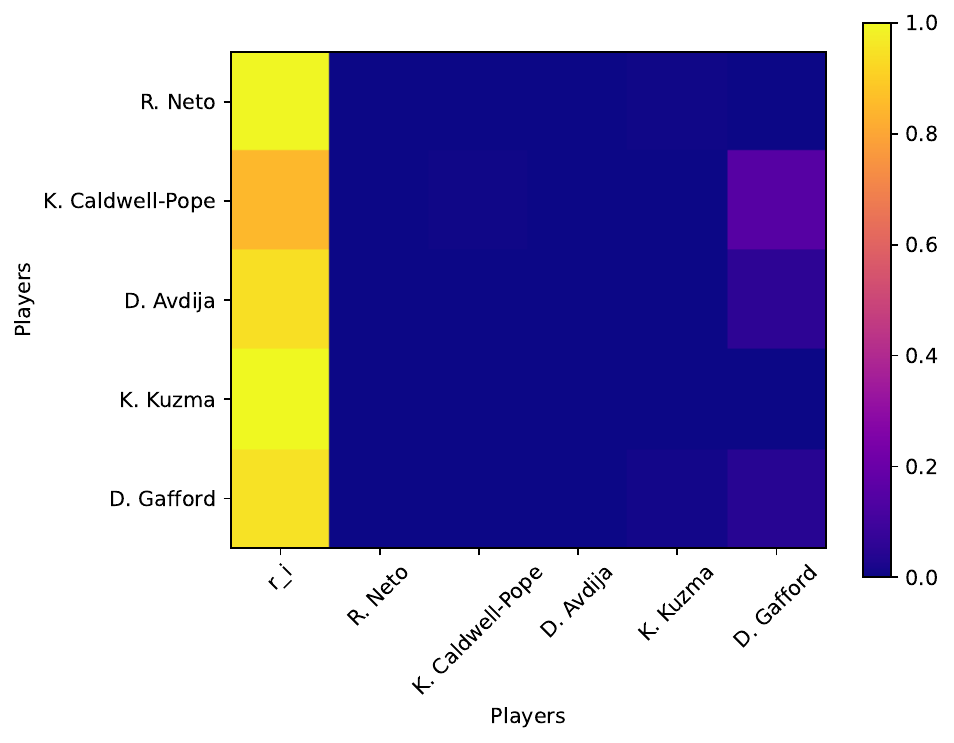}
        \caption{Washington Wizards}
    \end{subfigure}
    \hfill
    \begin{subfigure}[b]{\figwidth}
        \centering
    \includegraphics[width=\textwidth]{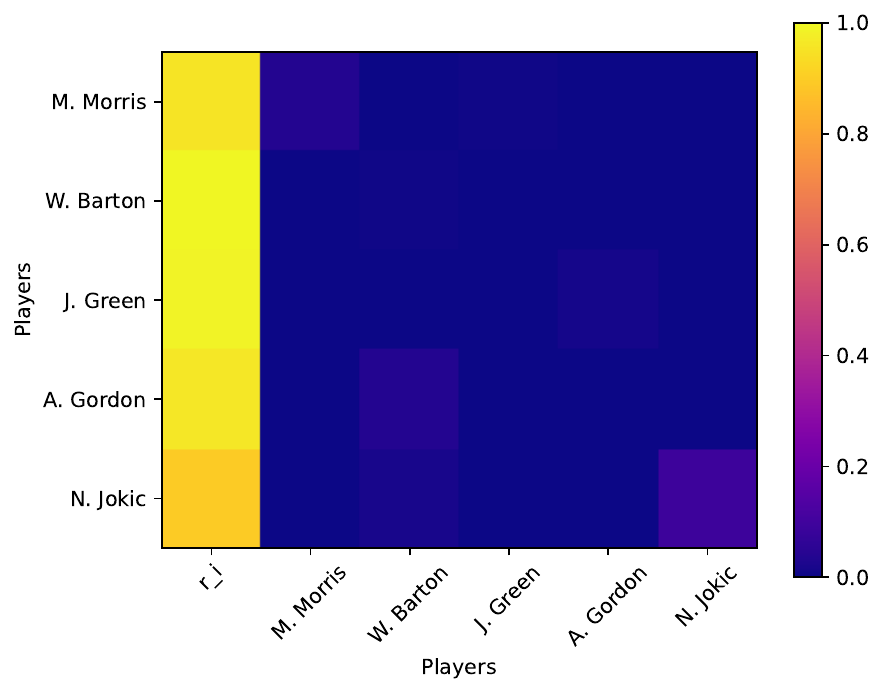}
        \caption{Denver Nuggets}
    \end{subfigure}
    \begin{subfigure}[b]{\figwidth}
        \centering
    \includegraphics[width=\textwidth]{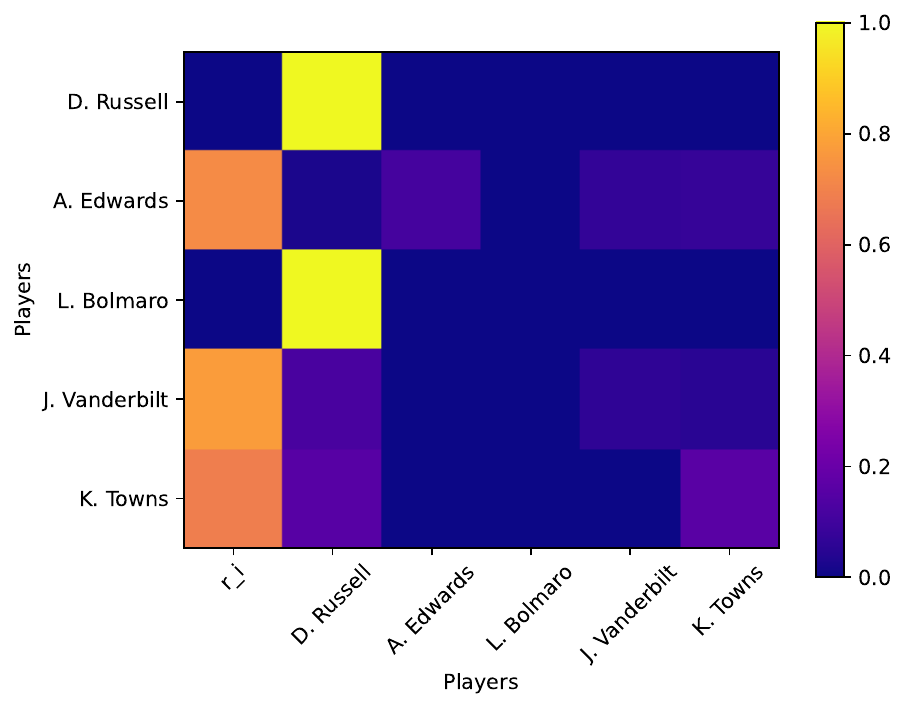}
        \caption{Minnesota Timberwolves}
    \end{subfigure}
    \hfill
    \begin{subfigure}[b]{\figwidth}
        \centering
    \includegraphics[width=\textwidth]{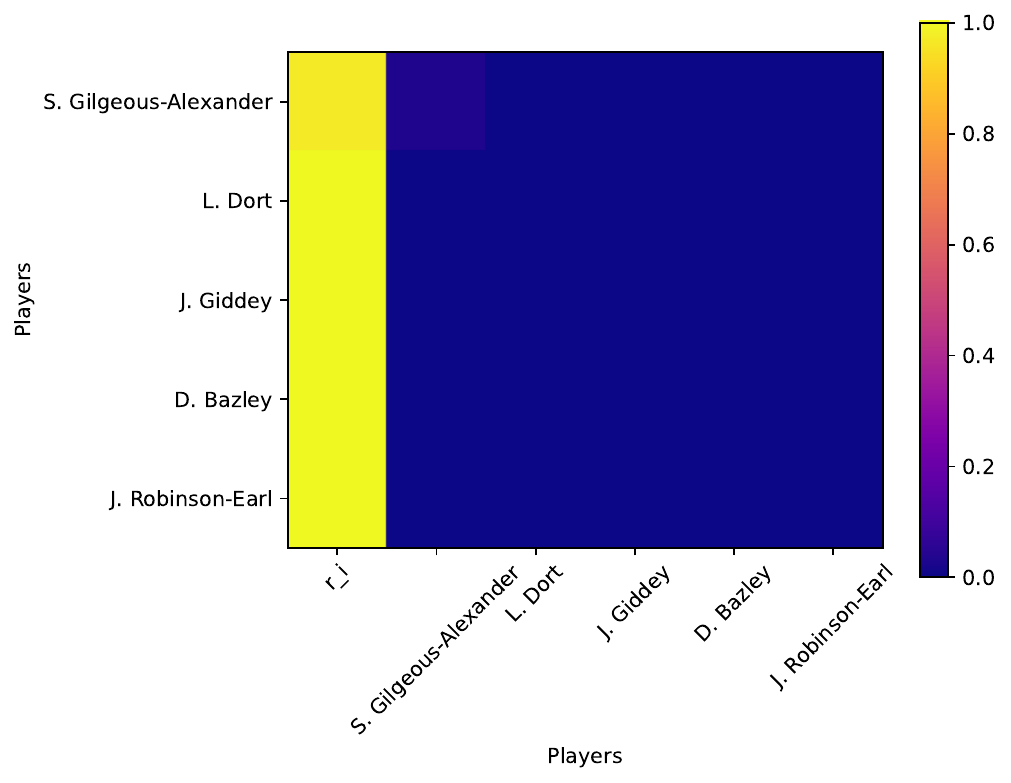}
        \caption{Oklahoma City Thunder}
    \end{subfigure}
    \caption{Team structure $\Rmatrix$}
    \label{fig: team-structure C}
\end{figure*}

\begin{figure*}[htbp]
    \centering
    \begin{subfigure}[b]{\figwidth}
        \centering
     \includegraphics[width=\textwidth]{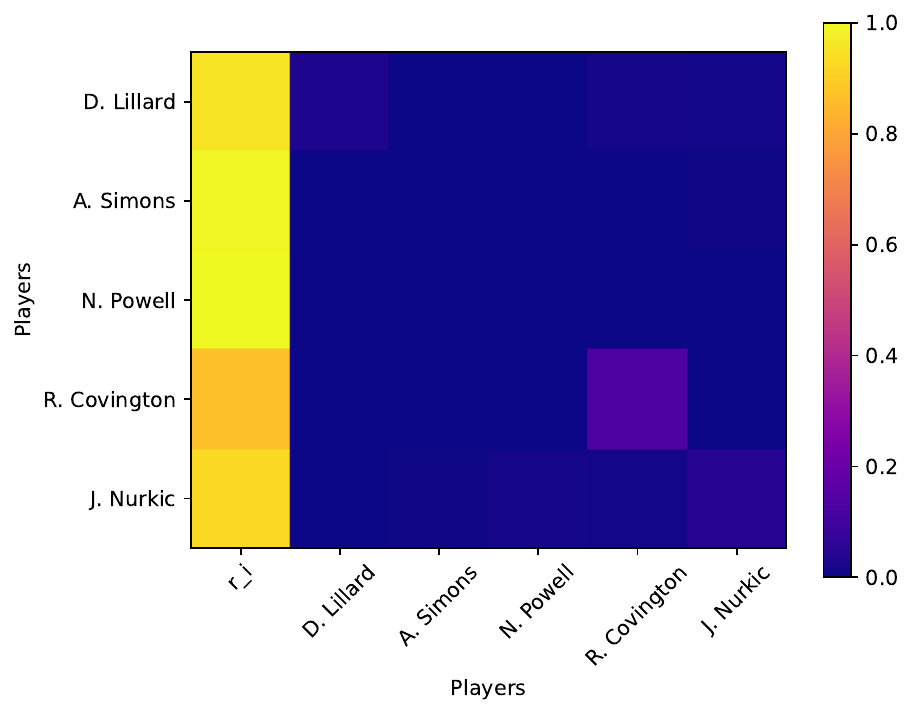}
        \caption{Portland Trail Blazers}
    \end{subfigure}
    \hfill
    \begin{subfigure}[b]{\figwidth}
        \centering
    \includegraphics[width=\textwidth]{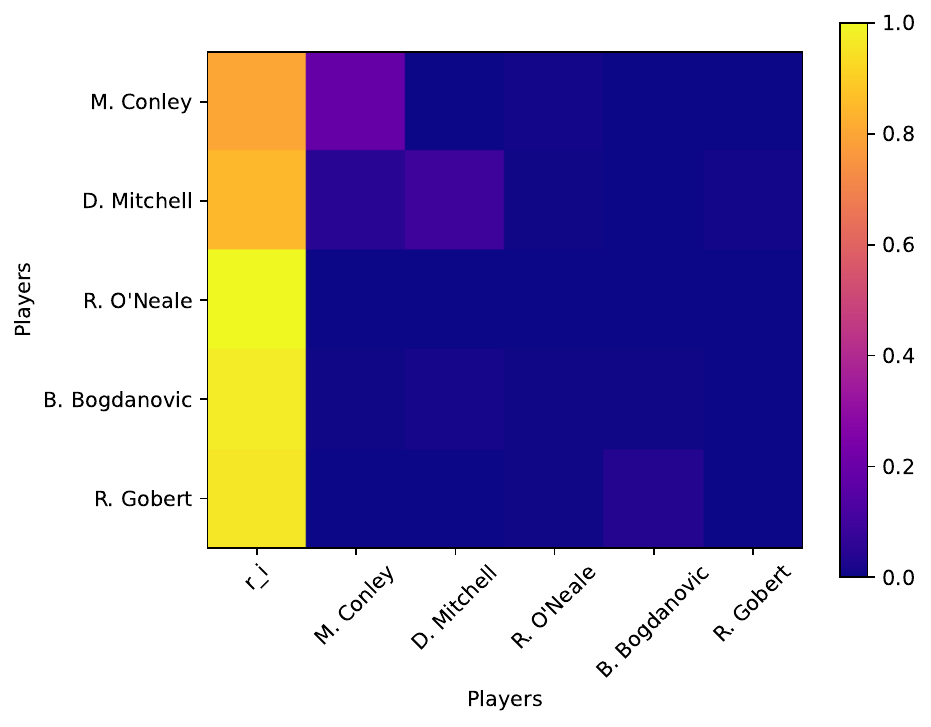}
        \caption{Utah Jazz}
    \end{subfigure}

    \begin{subfigure}[b]{\figwidth}
        \centering
    \includegraphics[width=\textwidth]{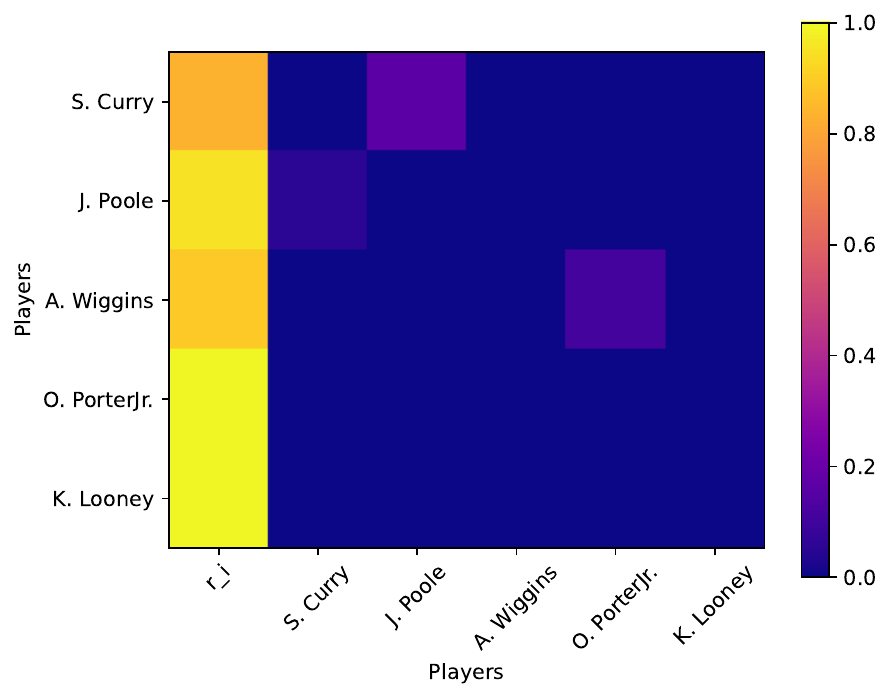}
        \caption{Golden State Warriors}
    \end{subfigure}
    \hfill
    \begin{subfigure}[b]{\figwidth}
        \centering
    \includegraphics[width=\textwidth]{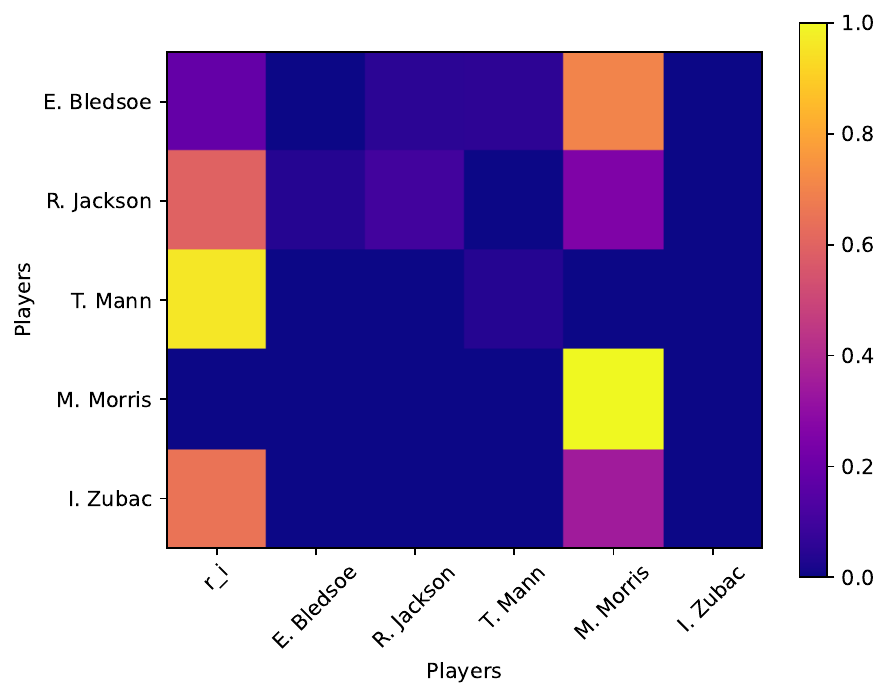}
        \caption{LA Clippers}
    \end{subfigure}
    \begin{subfigure}[b]{\figwidth}
        \centering
    \includegraphics[width=\textwidth]{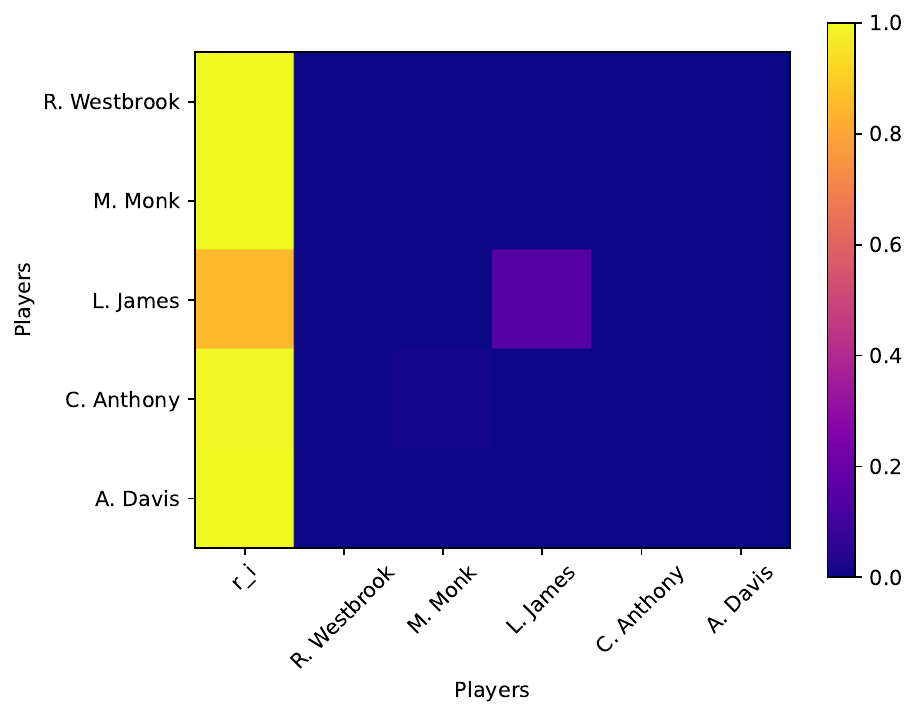}
        \caption{Los Angeles Lakers}
    \end{subfigure}
    \hfill
    \begin{subfigure}[b]{\figwidth}
        \centering
    \includegraphics[width=\textwidth]{img/heatmap/PHI_heatmap.pdf}
        \caption{Phoenix Suns}
    \end{subfigure}
    \caption{Team structure $\Rmatrix$}
    \label{fig: team-structure D}
\end{figure*}

\begin{figure*}[htbp]
    \centering
    \begin{subfigure}[b]{\figwidth}
        \centering
     \includegraphics[width=\textwidth]{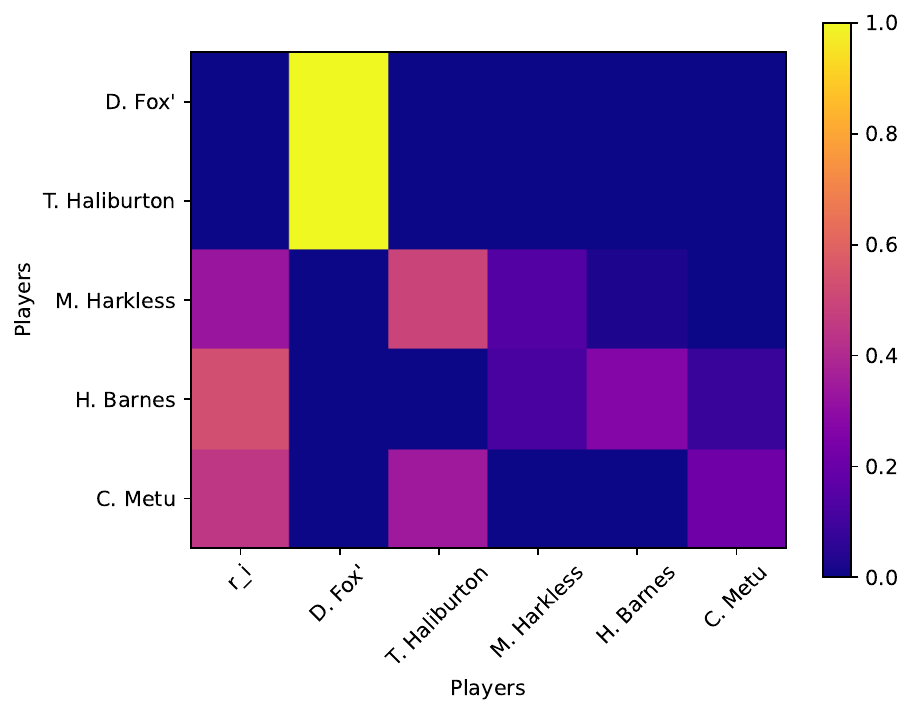}
        \caption{Sacramento Kings}
    \end{subfigure}
    \hfill
    \begin{subfigure}[b]{\figwidth}
        \centering
    \includegraphics[width=\textwidth]{img/heatmap/DAL_heatmap.pdf}
        \caption{Dallas Mavericks}
    \end{subfigure}

    \begin{subfigure}[b]{\figwidth}
        \centering
    \includegraphics[width=\textwidth]{img/heatmap/HOU_heatmap.pdf}
        \caption{Houston Rockets}
    \end{subfigure}
    \hfill
    \begin{subfigure}[b]{\figwidth}
        \centering
    \includegraphics[width=\textwidth]{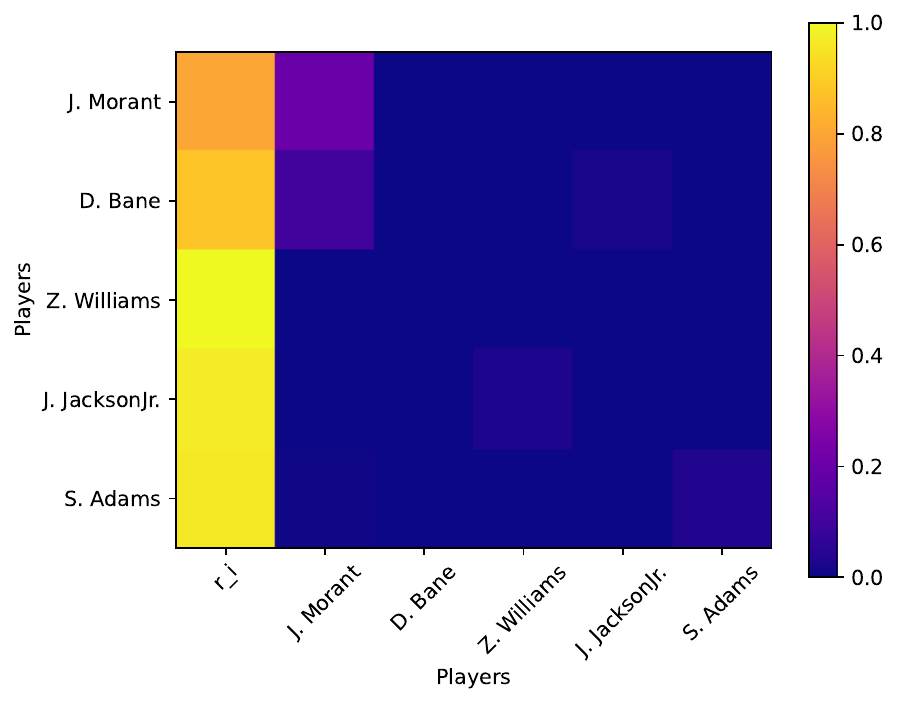}
        \caption{Memphis Grizzlies}
    \end{subfigure}
    \begin{subfigure}[b]{\figwidth}
        \centering
    \includegraphics[width=\textwidth]{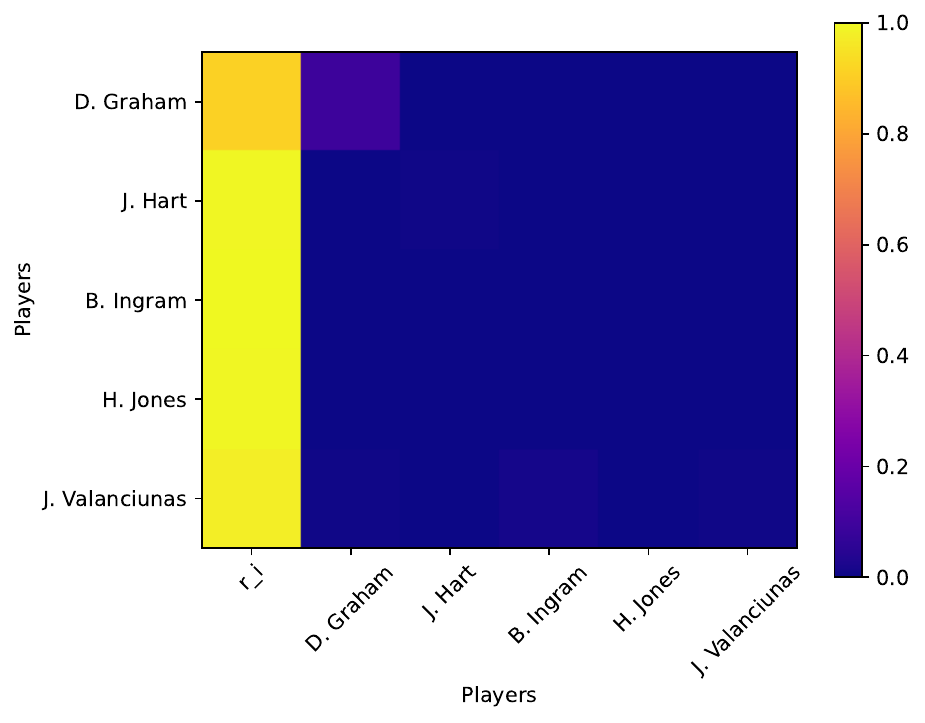}
        \caption{New Orleans Pelicans}
    \end{subfigure}
    \hfill
    \begin{subfigure}[b]{\figwidth}
        \centering
    \includegraphics[width=\textwidth]{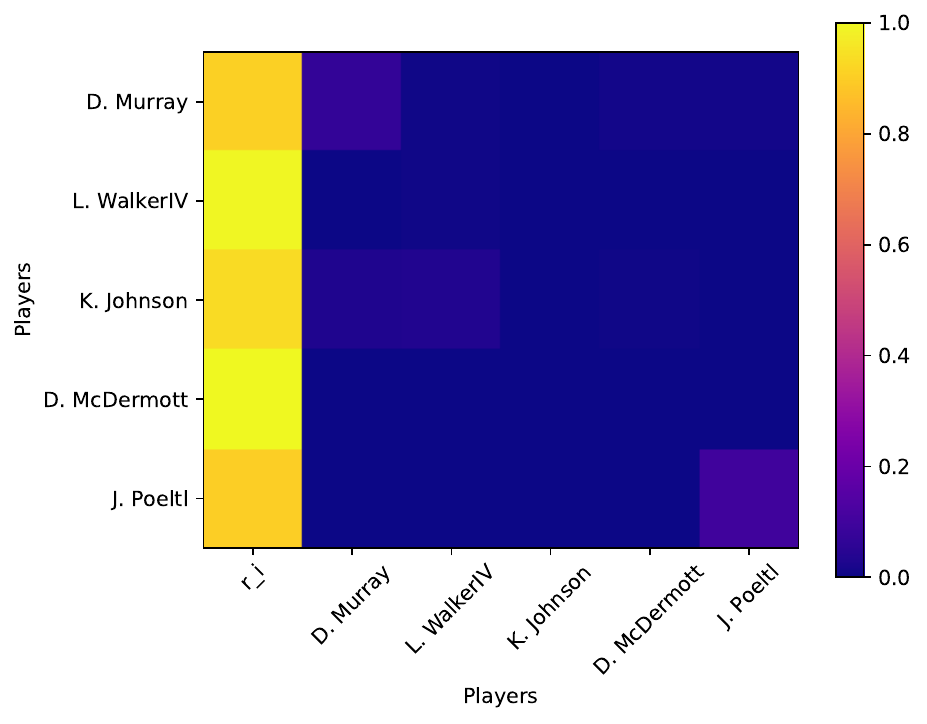}
        \caption{San Antonio Spurs}
    \end{subfigure}
    \caption{Team structure $\Rmatrix$}
    \label{fig: team-structure E}
\end{figure*}

\subsection{Additional examples of specific games} \label{app: NBA games}

The following figure provides some additional examples of games with the inferred from our model team state. 
As we can see here also, the team states inferred from our model follows closely in the vast majority of the cases the observed score differential at the game. 
All in all, the expressivity of our model is very good and can explain well the observed {\em output} of the game. 

\begin{figure*}[htbp]
    \centering
    \begin{subfigure}[b]{\figwidth}
        \centering
     \includegraphics[width=\textwidth]{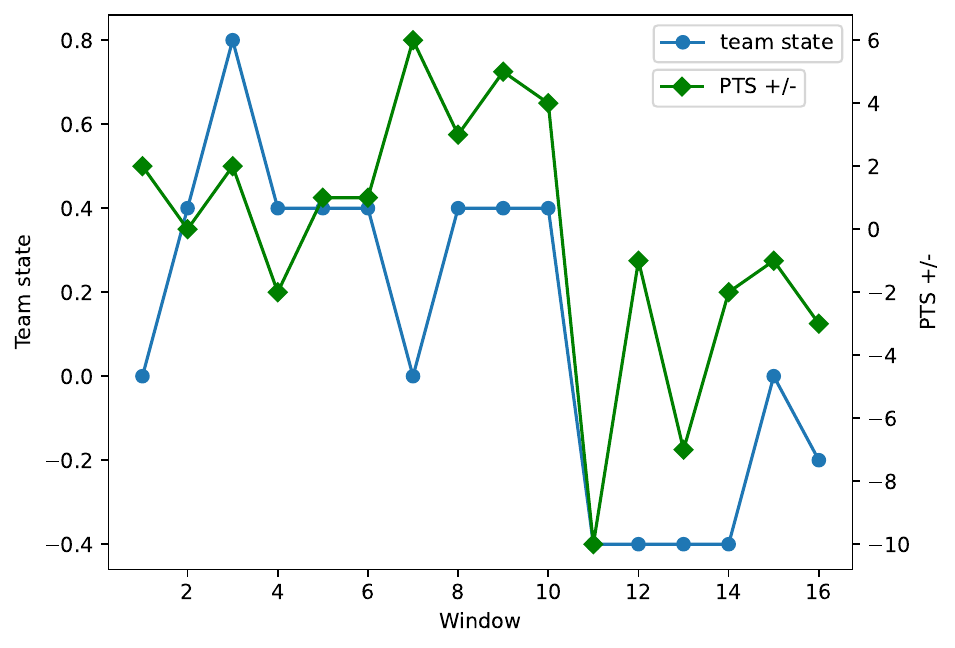}
        \caption{Denver Nuggets vs Los Angeles Lakers(11 Jan 2022)}
    \end{subfigure}
    \hfill
    \begin{subfigure}[b]{\figwidth}
        \centering
    \includegraphics[width=\textwidth]{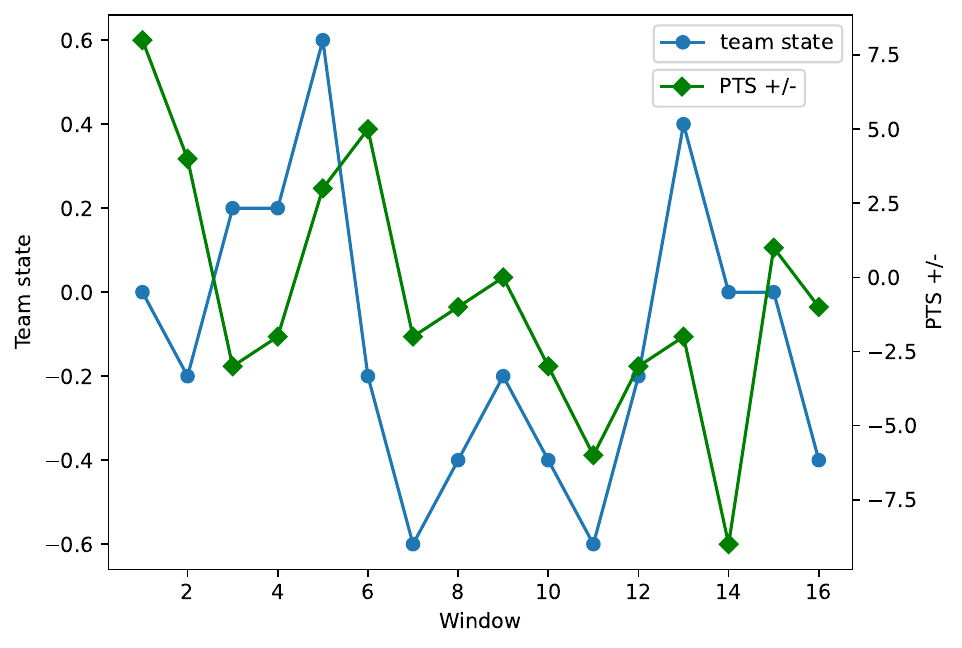}
        \caption{Dallas Mavericks vs Golden State Warriors (20 May 2022)}
    \end{subfigure}

    \begin{subfigure}[b]{\figwidth}
        \centering
    \includegraphics[width=\textwidth]{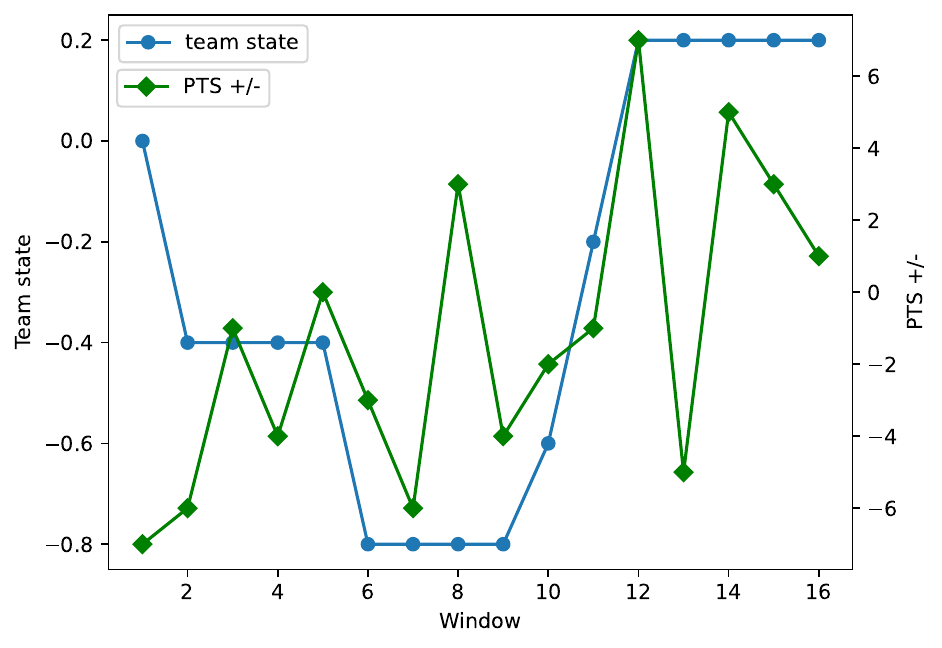}
        \caption{Miami Heat vs Boston Celtics (23 May 2022)}
    \end{subfigure}
    \hfill
    \begin{subfigure}[b]{\figwidth}
        \centering
    \includegraphics[width=\textwidth]{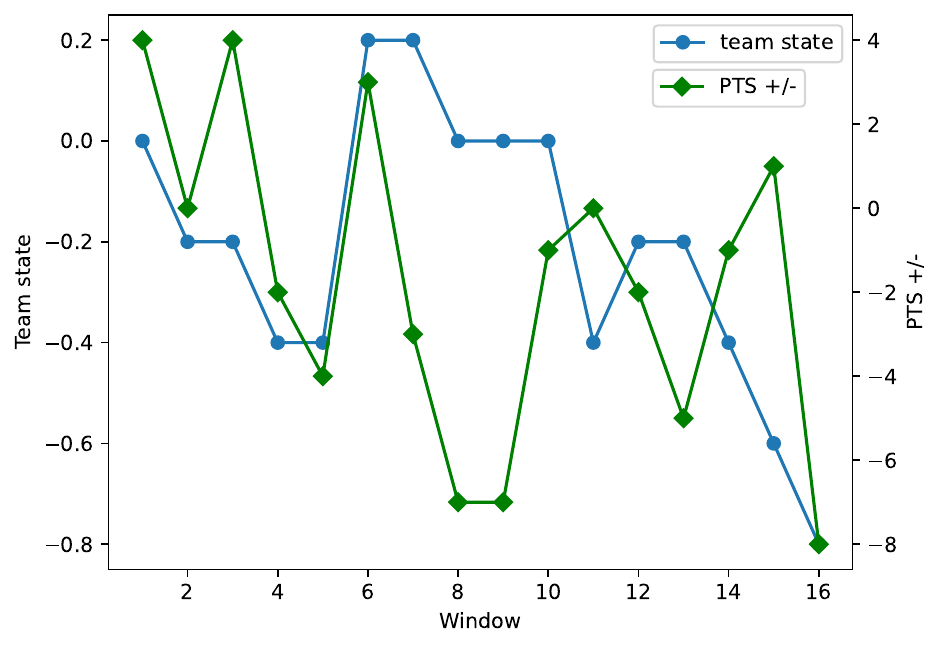}
        \caption{Milwaukee Bucks vs Boston Celtics (15 May 2022)}
    \end{subfigure}
    \caption{Team collapse games}
    \label{fig: NBA collapse games}
\end{figure*}